\documentclass[]{aa}
\usepackage{graphicx}
\usepackage{txfonts}

\def\uJAVA{\mbox{$u$}}
\def\gSDSS{\mbox{$g$}}
\def\rSDSS{\mbox{$r$}}
\def\iSDSS{\mbox{$i$}}
\def\zSDSS{\mbox{$z$}}
\def\Ja{\mbox{$J0378$}}
\def\Jb{\mbox{$J0395$}}
\def\Jc{\mbox{$J0410$}}
\def\Jd{\mbox{$J0430$}}
\def\Je{\mbox{$J0515$}}
\def\Jf{\mbox{$J0660$}}
\def\Jg{\mbox{$J0861$}}
\def\rT{\mbox{r$_{tidal}$}}
\def\rC{\mbox{r$_{core}$}}
\def\feH{\mbox{[Fe/H]}}
\def\mSun{\mbox{M$_\odot$}}
\def\lSun{\mbox{L$_\odot$}}

\begin{document} 

\title{J-PLUS: A wide-field multi-band study of the M\,15 globular cluster. Evidence of 
multiple stellar populations in the RGB.}

\author{Charles Bonatto\inst{1}, Ana L. Chies-Santos\inst{1}, Paula R. T. Coelho\inst{2}, Jes\'us 
Varela\inst{3}, S\o ren S. Larsen\inst{4}, A. Javier Cenarro\inst{3}, Izaskun San Roman\inst{3},
Antonio Mar\'\i n-Franch\inst{3}, Claudia Mendes de Oliveira\inst{2}, Alberto Molino\inst{2}, 
Alessandro Ederoclite\inst{3}, Arianna Cortesi\inst{2}, Carlos L\'opez-San Juan\inst{3}, David 
Crist\'obal-Hornillos\inst{3}, H\'ector V\'azquez Rami\'o\inst{3}, Laerte Sodr\'e Jr\inst{2}, Laura 
Sampedro\inst{2}, Marcus V. Costa-Duarte\inst{1}, Patr\'icia M. Novais\inst{2}, Renato Dupke\inst{5,6}, 
Roderik A. Overzier\inst{5}, Tiago Ribeiro\inst{7}, Walter A. Santos\inst{2}, \and William Schoennell\inst{2,8}   } 
         
\offprints{C. Bonatto}

\institute{Departamento de Astronomia, Instituto de F\'isica, Universidade Federal 
do Rio Grande do Sul, Porto Alegre, RS, Brazil\\
   \email{charles.bonatto,ana.chies@ufrgs.br}
   \and Universidade de S\~ao Paulo, S\~ao Paulo, Instituto de Astronomia, Geof\'isica e Ci\^encias 
   Atmosf\'ericas, SP, Brazil
   \and
   Centro de Estudios de F\'isica del Cosmos de Arag\'on, Plaza San Juan 1, 44001 Teruel, Spain. 
   \and
    Department of Astrophysics/IMAPP, Radboud University, Nijmegen, The Netherlands
    \and
    Observat\'orio Nacional, RJ, Brazil
    \and
     University of Michigan, Ann Arbor, US
    \and
     Departamento de F\'isica, Universidade Federal de Sergipe, Av. Marechal Rondon s/n, 49100-000 
     S\~ao Crist\'ov\~ao, SE, Brazil
    \and
      AA-CSIC, Glorieta de la Astronom\'ia S/N. E-18008, Granada, Spain  } 
\date{}
 
  \abstract
   {As a consequence of internal and external dynamical processes, Galactic globular clusters (GCs) 
have properties that vary radially. Wide-field observations covering the entire projected area of GCs 
out to their tidal radii (\rT) can therefore give crucial information on these important relics of the 
Milky Way formation era. }
   {The Javalambre Photometric Local Universe Survey (J-PLUS) provides wide field-of-view ($2\,deg^2$) 
images in 12 narrow, intermediate and broad-band filters optimized for stellar photometry. Here we have applied 
J-PLUS data for the first time for the study of Galactic GCs using science verification data obtained for 
the very metal-poor ($\feH\approx-2.3$) GC M\,15 located at $\sim10$\,kpc from the Sun. Previous studies 
based on spectroscopy found evidence of multiple stellar populations (MPs) through their different 
abundances of C, N, O, and Na. Our J-PLUS data provide low-resolution spectral energy distributions 
covering the near-UV to the near-IR, allowing us to instead search for MPs based on pseudo-spectral 
fitting diagnostics. }
   {We have built and discussed the stellar radial density profile (RDP) and surface 
brightness profiles (SBPs) reaching up to \rT. Since J-PLUS FoV is larger than M\,15's \rT, the field
contamination can be properly taken into account. We also demonstrated the power of J-PLUS unique filter 
system by showing colour-magnitude diagrams (CMDs) using different filter combinations and for different 
cluster regions.}
  {J-PLUS photometric quality and depth are good enough to reach the upper end of M\,15's main-sequence. 
CMDs based on the colours $(\uJAVA-\zSDSS)$ and $(\Ja-\Jg)$ are found to be particularly useful to 
search for splits in the sequences formed by the upper red giant branch (RGB) and asymptotic giant branch 
(AGB) stars. We interpret these split sequences as evidence for the presence of MPs. Furthermore, we show
that the $(\uJAVA-\zSDSS)\times (\Ja-\gSDSS)$ colour-colour diagram allows us to distinguish clearly between 
field and M\,15 stars, which is important to minimize the sample contamination. }
  {The J-PLUS filter combinations $(\uJAVA - \zSDSS)$ and $(\Ja - \Jg)$, which are sensitive to metal 
abundances, are able to distinguish different sequences in the upper RGB and AGB regions of the CMD of 
M\,15, showing the feasibility of identifying MPs without the need of spectroscopy. This demonstrates 
that the J-PLUS survey will have sufficient spatial coverage and spectral resolution to perform a large 
statistical study of GCs through multi-band photometry in the coming years. }

\keywords{({\em Galaxy}:) globular clusters: general; ({\em Galaxy}:) globular clusters: individual: 
M\,15; surveys}
  
\titlerunning{J-PLUS study of M\,15}
   
\authorrunning{Bonatto et al.}

\maketitle
%

\section{Introduction}
\label{Intro}

Galactic globular clusters (GCs) are among the most interesting - and beautiful - relics of the Milky Way 
formation epoch. Given their long-lived nature, some intrinsic properties (e.g. age, metallicity, and mass
distribution) and the large-scale spatial distribution of GCs formed in the early phases of the Galaxy may
provide clues to the Milky Way assembly process. In this sense, studies of GC properties can be used to set
constraints to Galaxy formation models, as well as stellar and dynamical evolution theories.

In general, GCs are relatively isolated, self-gravitating, and dynamically relaxed (e.g.
\citealt{Dabringhausen2008}) multi-particle systems. There is scarce evidence of the presence of 
dark matter in GCs, which is consistent with expectations for small dark-matter halos and, besides, 
they have low mass-to-light ratios ($M/L_V\approx0.5 - 3.5\,\mSun/\lSun$), probably because of 
the preferential loss of low-mass stars related to their advanced dynamical evolution (e.g. 
\citealt{Dabringhausen2008}).

\begin{figure*}
\resizebox{\hsize}{!}{\includegraphics{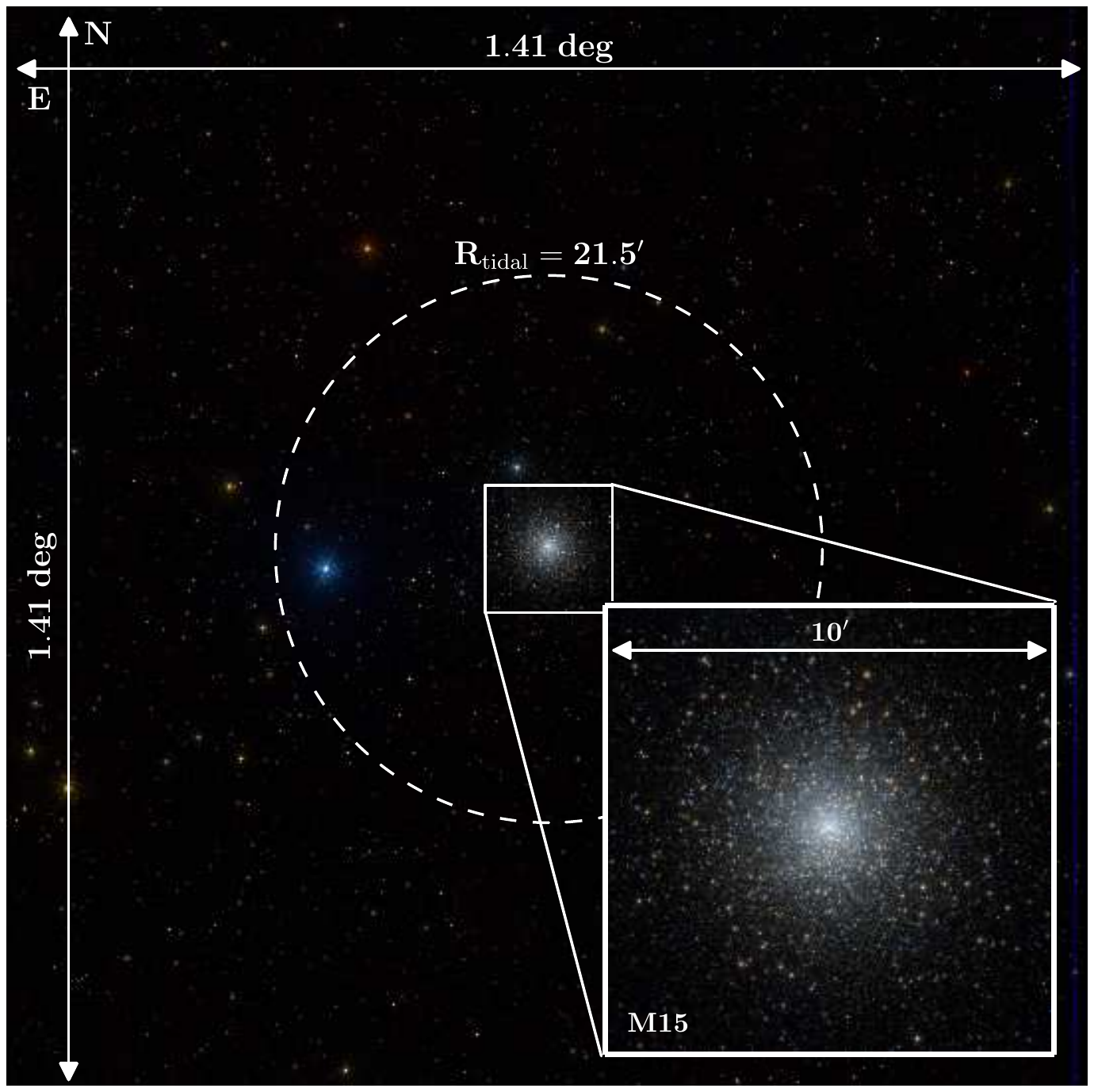}}
\caption[M\,15 globular cluster]{Composite three-colour image of M\,15 built with the 
J-PLUS filters \Je, \Jf\ and \Jg; the tidal radius ($\rT\sim21.5\arcmin$) is fully encompassed 
by the observation, which shows the complete FoV of the instrument. Inset: magnification of the brightest 
part of M\,15 ($R\sim5\arcmin$). }
\label{fig1}
\end{figure*} 

Once considered as the prototypes of simple (or single) stellar populations and the building blocks of galaxies, 
GCs now have to be described by more complex - and theoretically challenging - scenarios, as shown by a growing 
number of works being published since the past decade (for recent reviews, see \citealt{Gratton2012} and 
\citealt{Bastian2017}). More specifically, high-quality spectroscopy (e.g. \citealt{Carretta2015}) and precision 
photometry (e.g. \citealt{Piotto2007}) led to the unambiguous detection of multiple stellar populations (MPs) 
in several Milky Way GCs as well as in old and young, massive extragalactic star clusters, like in the Magellanic 
Clouds (\citealt{Milone2017}), although it's not clear if the processes acting on the young Magellanic Clouds clusters
are the same at work in the old ones. The presence of MPs is so ubiquitous that it became a rule for high-mass 
($M>2\times10^5\,\mSun$) GCs (\citealt{Gratton2012}). However, we remark that multiple populations have also 
been detected further down in the Galactic GC mass function, as are the recently reported cases of NGC\,6362 
(\citealt{Dalessandro2014}) and NGC\,6535 (\citealt{Piotto2015}). 

Spectroscopy is the most direct way to find systematic abundance variations among groups of stars and, thus, 
physically characterize MPs, but it is usually restricted to relatively small stellar samples. Photometry, on 
the other hand, has been shown to be an excellent alternative tool with which to search for MPs by tracking 
separate evolutionary sequences - in colour-magnitude diagrams (CMDs) of GCs - occupied by many stars. For its
part, this technique relies on relatively high-precision photometry and, as such, Hubble Space Telescope (HST) 
data has helped revealing such features in different regions of many GCs (e.g. \citealt{Piotto2007}). However, 
most of the current spectroscopy and imaging on GCs has been based on data from instruments covering 
partially or even small fractions of the GC body, such as ACS, WFPC2 and WFPC3 (e.g. \citealt{Milone2009},
\citealt{Milone2013}) of HST, as well as FLAMES and FORS2 (e.g. \citealt{Gratton2010}, \citealt{Nardiello2015}) 
at the Very-Large Telescope (VLT/ESO). In fact, with a FoV of $\sim25\arcmin$ in diameter, ESO/FLAMES can sample 
from the very centre to the \rT\ of about half of the Galactic GCs listed in \citet{Harris2010}.

The {\em Javalambre Photometric Local Universe Survey} (J-PLUS)\footnote{J-PLUS website: http://www.j-plus.es/} 
will observe $8500\deg^2$ of the sky visible from the Javalambre Observatory in Spain with the panoramic camera 
T80Cam at the JAST/T80 telescope using a set of 12 broad, intermediate and narrow-band optical filters. Some of 
these filters are located at key stellar spectral features (see Table~\ref{tab1}).

The wide-field ($1.4\degr\times1.4\degr$) capabilities of T80Cam may be an excellent tool to study integrated 
properties of Galactic GCs, such as mass and luminosity segregations, total mass, among others. Coupled to the 
set of 12 optimized narrow, intermediate, and broad-band filters - providing adequate sampling of the stellar 
spectral energy distribution (SED) - we will have tools to explore specific pseudo-spectral fitting diagnostics 
to photometrically infer about the presence of MPs. In addition, the data, covering all - or most - of the GC body, 
can also be used to search for MPs in different regions of the clusters. As a test case, we have obtained 
science verification data in the 12 J-PLUS bands for the GC M\,15 (NGC\,7078). M\,15 is old ($\ga$10\,Gyr), 
and very metal-poor $\feH=-2.3$ \citep{Carretta2009}, located $\sim$10\,kpc away from the Sun and 10\,kpc 
away from the Galactic centre \citep{Harris2010}. The core and tidal radii of M\,15 are \citep{Harris2010} 
$\rC=0.14\arcmin$ ($\sim0.4$\,pc) and $\rT=21.5\arcmin$ ($\sim60$\,pc), respectively. It is a post-core-collapsed 
GC (\citealt{DjorKing1986}; \citealt{MarchiParesce96}) where the presence of a central intermediate-mass black hole 
is still under debate (e.g. \citealt{Kirsten2014}).

MPs in M\,15 have been detected both spectroscopically (e.g. \citealt{Carretta2009}) and photometrically 
(e.g. \citealt{Lardo2011}; \citealt{Piotto2015}). In particular, \citet{Larsen2015} employed photometry from 
HST/WFC3 and Sloan Digital Sky survey (SDSS) to confirm the presence of MPs (at least in the radial range 
$0.07\arcmin - 2.2\arcmin$ around the GC centre) in the lower red-giant branch (RGB), which is indicative of 
1st and 2nd generation populations dominating at different cluster-centric distances. They found radial variations 
in possibly 3 populations of stars characterized by different abundances of C, N, O, and Na.

In this paper we have studied the stellar radial density and surface brightness profiles (RDP and SBP) for the 12 J-PLUS 
bands, covering the radial range from the centre to the tidal radius of M\,15. When reaching the full radial range 
of a given GC, both kinds of profiles are important to study the large-scale structure and might provide crucial 
clues on the dynamical state. We have also investigated the evolutionary sequences of M\,15 in different colour-magnitude 
diagrams, and tried to detect MPs in its giant branches.

This paper is organized as follows: in Sect.~\ref{Data} we briefly discuss the instrumentation and 
observations, and present the photometry. In Sect.~\ref{Analysis} we show the spatial distribution
of stars and surface brightness, in Sect.~\ref{JPLUS_CMD} we discuss properties 
of colour-magnitude diagrams. Finally, discussions and our conclusions are given in Sect.~\ref{conclu}.

\section{Instrumentation and data}
\label{Data}

The J-PLUS survey is being carried out with the Javalambre Auxiliary Survey Telescope (JAST/T80), an 0.83 meter, 
F/4.5 telescope with a German-equatorial mount and a Ritchey-Chr\'etien optical configuration. It is installed
at the Observatorio Astrof\'\i sico de Javalambre (OAJ) in Spain. The focal plane corresponds to a Cassegrain 
layout with a field corrector of three spherical lenses of fused silica in the range $115-140$\,mm. The JAST/T80 
field of view (FoV) has a diameter of $1.7\degr$ (110\,mm), with full optical performance, reaching the 
$2.0\degr$ (130\,mm) diameter with a slight vignetting ($<1\%$) and slightly degraded image quality. The 
scientific instrument for the JAST/T80 telescope (and for J-PLUS) is the imager T80Cam, a wide-field camera 
installed at the Cassegrain focus. It is equipped with a large-format $9232\times9216$ pixels e2v CCD 
($10\,\mu \rm m$ pixel), with a $0.55\arcsec\,{\rm pixel}^{-1}$ scale yielding a FoV of $1.4\degr\times1.4\degr$, 
and a set of 12 broad, intermediate and narrow band filters. The high-efficiency CCD is read from 16 ports 
simultaneously, allowing readout times of 12s with a typical readout noise of 3.4 electrons (RMS). 

The J-PLUS survey has been primarily conceived to perform the calibration for the main J-PAS\footnote{J-PAS website:
http://www.j-pas.org/} survey data, and will observe more than $8500\deg^2$ of the sky through a set of 
12 filters. The photometric system of the survey is composed of 4 broad (\gSDSS, \rSDSS, \iSDSS\ \& \zSDSS), 
two intermediate (\uJAVA\ \& \Jg) and six narrow-band (\Ja, \Jb, \Jc, \Jd, \Je\ \& \Jf) filters covering the 
whole optical range. This is an optimized filter system to properly recover the stellar parameters 
$T_{eff}$, $\log(g)$ and \feH, through the fitting of flux calibrated models of the observed stars. 
Further details are in \citet{cenarro18}, hereafter referred to as Paper\,0. Concerning the {\em Javalambre 
Physics of the Accelerating Universe Astrophysical Survey} (J-PAS), this survey will cover at least $8000\deg^2$ 
in approximately five years, using an unprecedented system of 56 narrow band filters in the optical. The main J-PAS 
instrument is a 2.5\,m telescope with an effective FoV of $5\deg^2$. Besides photometry, J-PAS will deliver
high-quality redshift and low-resolution spectroscopy for hundreds of millions of galaxies. In addition to the 
present paper, the J-PLUS EDR and science verification data were used to
refine the membership in nearby galaxy clusters (\citealt{molino18}), study the H$\alpha$
emission (\citealt{logronho18}) and the stellar populations (\citealt{sanroman18}) of several
local galaxies, and compute the stellar and galaxy number counts up to $r = 21$ (\citealt{clsj18}).

\begin{table}
\caption{{\small Informations on the observations of M\,15}}
\label{tab1}
\renewcommand{\tabcolsep}{0.7mm}
\begin{tabular}{lcccccccccccccccccc}
\hline
\hline
Filter & $\lambda_{eff}$ & $\Delta \lambda$  & $t_{sng}$ & Total & Seeing & N & m$_{lim}$ & Spectral\\
name & [nm] & [nm] & [sec] & [sec] & [\arcsec] & (stars) & (mag) & Feature\\
\hline
(1) & (2) & (3) & (4) & (5) & (6) & (7) & (8) & (9) \\
\hline
\uJAVA    &  348.5 &  50.8  &  60  &  600  & 1.944 & 13451 & 22.2 & \\
\Ja       &  378.5 &  16.8  &  60  &  544  & 1.873 &  6289 & 20.2 & [OII]\\
\Jb       &  395.0 &  10.0  &  60  &  604  & 1.643 &  7532 & 19.2 & Ca H$+$K\\
\Jc       &  410.0 &  20.0  &  60  &  605  & 1.473 & 22522 & 21.6 & H$_\delta$\\
\Jd       &  430.0 &  20.0  &  60  &  605  & 1.481 & 22959 & 21.7 & G-band\\ 
\gSDSS    &  480.3 & 140.9  &  10  &  115  & 1.383 & 39427 & 22.6 & \\
\Je       &  515.0 &  20.0  &  60  &  665  & 1.299 & 23968 & 21.6 & Mgb\\
\rSDSS    &  625.4 & 138.8  &  10  &  105  & 1.179 & 53357 & 22.4 & \\
\Jf       &  660.0 &  13.8  &  60  &  605  & 1.214 & 22402 & 20.6 & H$_\alpha$\\
\iSDSS    &  766.8 & 153.5  &  10  &  105  & 1.117 & 57518 & 21.8 & \\
\Jg       &  861.0 &  40.0  &  60  &  675  & 1.211 & 40723 & 21.0 & Ca-Trpl\\
\zSDSS    &  911.4 & 140.9  &  10  &  105  & 1.071 & 49510 & 21.2 & \\
\hline
\end{tabular}
\begin{list}{Table Notes.}
\item Col.~2: central wavelength of the filter; Col.~3: effective pass band; Col.~4: exposure time 
of individual  frames.; col.~5: total exposure time of the co-added image; Col.~6: average seeing of 
the short-exposure frames; Col.~7: number of stars with error $\le0.5$\,mag; Col.~8: limiting (faintest) 
magnitude reached with error $\le0.5$\,mag; Col.~9: key spectral feature sampled.
\end{list}
\end{table}

\begin{figure}
\resizebox{\hsize}{!}{\includegraphics{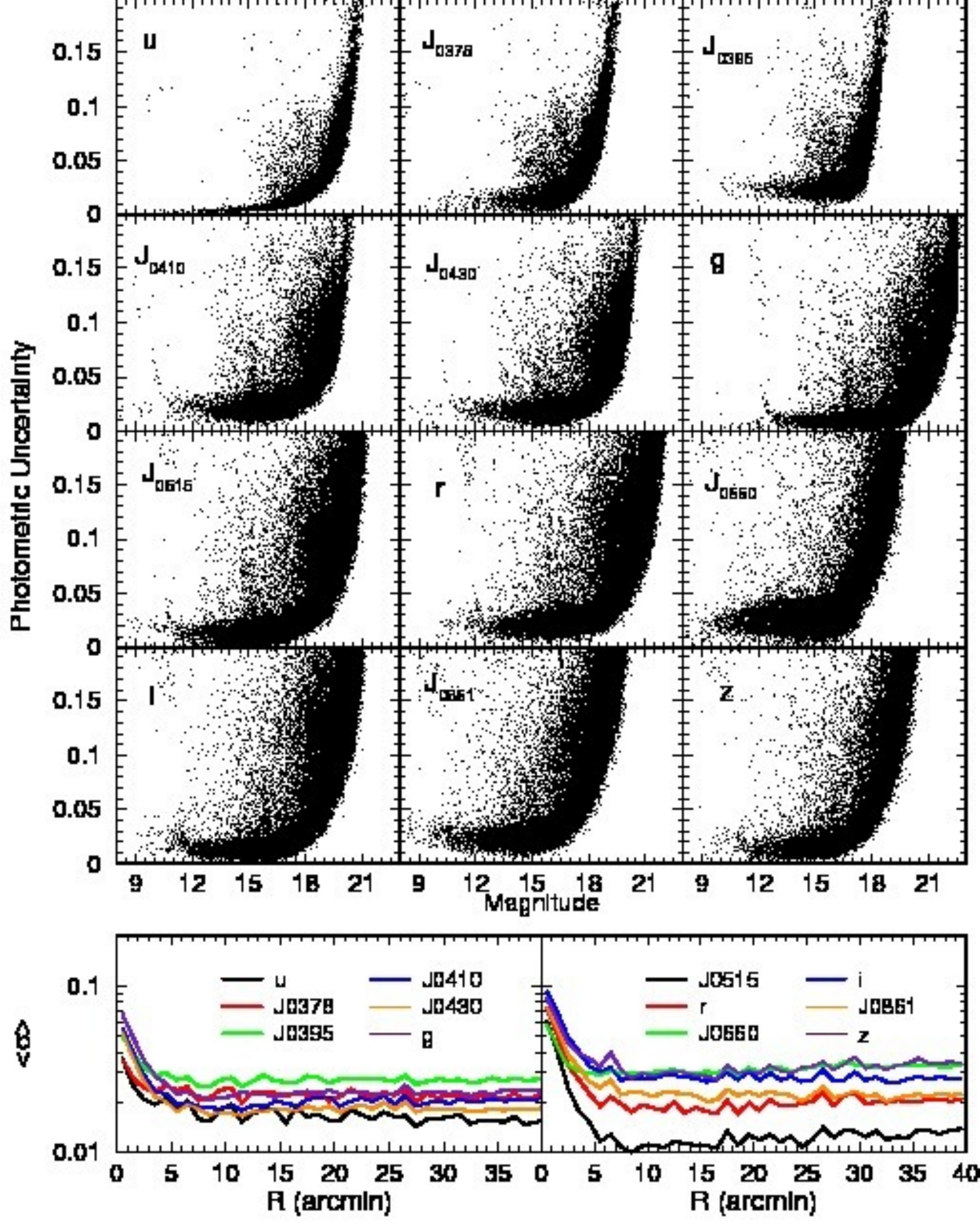}}
\caption[]{Top panels: Photometric error as a function of the measured magnitude. Bottom: Average 
photometric error as a function of the distance to M\,15 centre. Filters are indicated in panels. }
\label{fig2}
\end{figure} 

In this paper we present science verification observations\footnote{Already be made public at 
early-data release at the J-PLUS web portal: http://j-plus.es/datareleases/early\_data\_release.} of the Galactic GC M\,15, collected during the nights of 13 and 14 of November 
2015. Table~\ref{tab1} summarizes the journal of the observations, provides basic information 
on the filters and gives some statistics on the photometric catalogues. Given the wide T80Cam FoV, a 
single pointing was more than enough to cover the entire projected area of M\,15. At a distance from 
the Sun d$_\odot\approx10$\,kpc, 1\arcmin\ corresponds to $\approx2.9$\,pc. To avoid saturation of 
bright RGB and AGB stars, the final co-added images consist of a relatively large number of 
short-exposure individual frames, with a minimum number of ten images. We remark that these science 
verification images are significantly deeper than those obtained with the adopted exposure-time for 
the standard J-PLUS strategy (see Paper\,0). The number of stars and the limiting (faintest) magnitude 
detected with uncertainty $<0.5$\, mag in each image are also given in Table~\ref{tab1} for each filter.

Data reduction, photometric zero-points and co-added images for each filter were provided by OAJ/CEFCA 
(see Paper\,0 for details). We have built photometric catalogues for all the J-PLUS filters by performing 
DAOPHOT PSF (\citealt{Stetson1987}) photometry to each co-added image. The standard procedure of OAJ is
to use SExtractor (\citealt{Bertin96}) to extract photometry for the stars present in a given image. 
However, for crowded fields such as those containing Galactic GCs, PSF DAOPHOT photometry reaches deeper 
photometric limits than SExtractor. In the present case, we reached limits of one to two magnitudes deeper than 
the standard J-PLUS SExtractor photometry. To build the PSF model, we have used stars distributed all
over the co-added image of a given filter. Figure~\ref{fig1} is a three-colour composite image of M\,15, 
which illustrates the wide-field capabilities of the instrument, reaching beyond the tidal radius of 
the cluster ($\rT\sim21.5\arcmin\sim62$\,pc) with uniform photometry. We note that the K\,648 planetary 
nebula (\citealt{Kustner21}, \citealt{Pease28}) appears as a greenish point located 
$\approx30\arcsec$ ($\approx1.4$\,pc) to the north-east of M\,15 centre.

The quality of the photometry in each filter can be assessed by the dependence of the uncertainty 
($1\,\sigma$) on the measured magnitude. These relations are shown in Fig.~\ref{fig2} for all
stars contained in the final reduced images, whose panels contain only stars with uncertainties lower 
than $0.2$\,mag. In most cases, the error {\em vs} magnitude dependence follows the usual exponential 
increase after some magnitude. Most importantly, the uncertainty for stars in the RGB, horizontal branch 
(HB) and AGB is lower than 0.05\,mag in all filters. Such small errors indicate that photometric scattering 
should not be significant along the evolutionary sequences occupied by these stars, as it will be shown 
later. Based on the above, we remark that only stars with photometric uncertainty <$0.2$\,mag will be used 
in the remaining analysis. In addition, we also examine in the bottom panels of Fig.~\ref{fig2} the 
dependence of the average error (<$\sigma$>) on the distance to M\,15 centre. As expected from crowding, 
there is a slight increase of <$\sigma$> for $R\la5\,\arcmin$ ($\approx15$\,pc) - especially for the filters 
redder than \Je, but hardly reaching 0.1\,mag.

\section{Full-scale radial analysis of M\,15}
\label{Analysis}

Globular clusters have long been considered as excellent laboratories for the study of stellar 
dynamical processes (e.g. \citealt{Noyola2007}) and N-body simulations (e.g. \citealt{Baumgardt2017},
\citealt{Khalisi2007}). Internal dynamical processes such as core-collapse, the presence of a central 
black hole, as well as the physics of the initial gravitational collapse will affect the central region 
in different levels, while external tidal stresses plus evaporation and ejection are expected to produce 
observable effects at larger radii (e.g. \citealt{Noyola2007}). These effects are expected to leave
observable imprints on large-scale radial profiles. In this sense, it is desirable to have 
access to (uniform and relatively deep) photometric data that cover all - or most - of the GC body.
In this section we discuss the radial behaviour of some properties obtained from our 
wide-field observations of M\,15.

\subsection{Stellar radial density distribution}
\label{s_RDP}

\begin{figure}
\resizebox{\hsize}{!}{\includegraphics{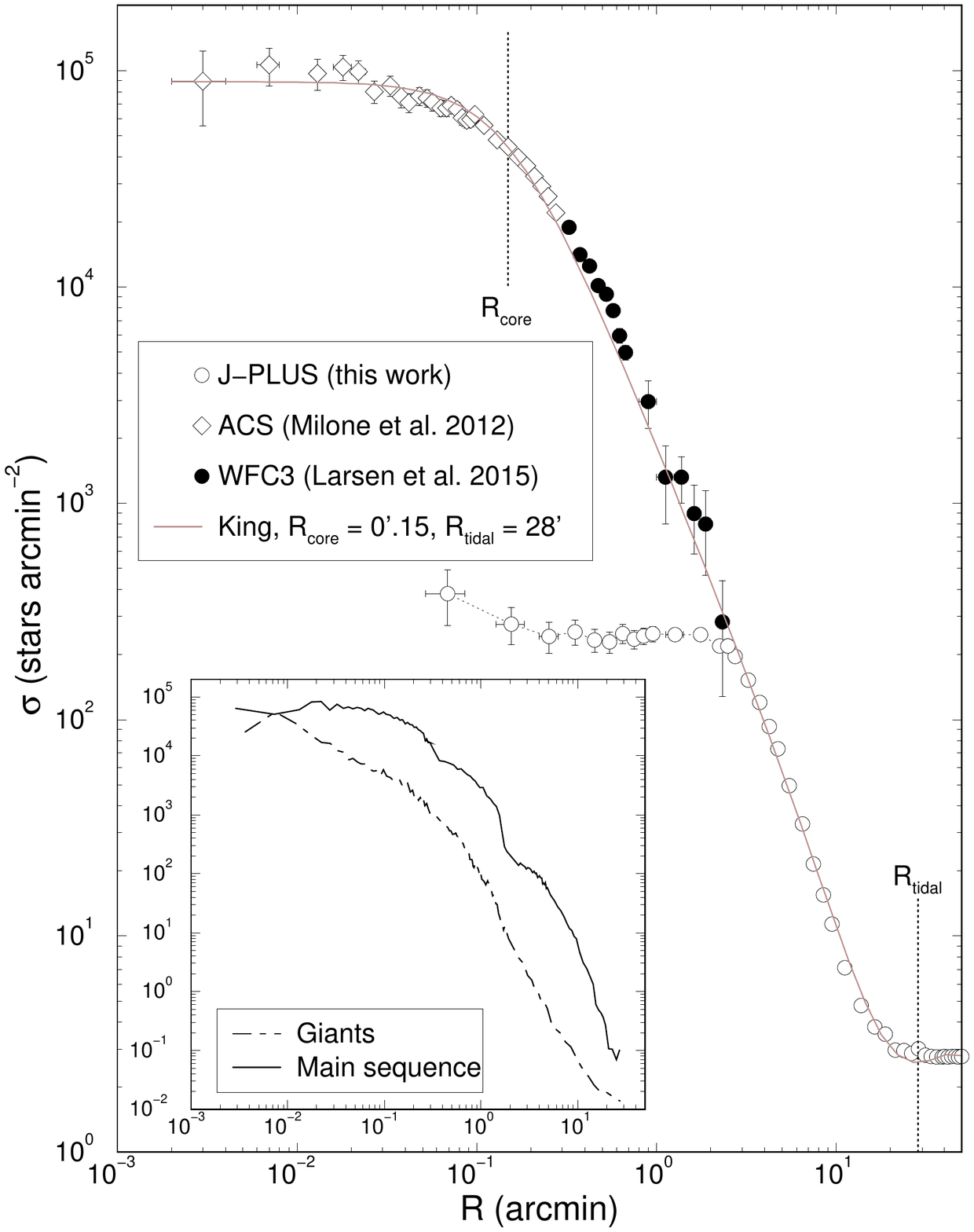}}
\caption[]{Composite stellar radial density profile of M\,15. HST data has been used for the region
where J-PLUS suffers severe incompleteness ($R\le3\arcmin$). Core and tidal radii derived from fitting 
the 3-par King profile are indicated. Inset: RDPs for main-sequence (solid line) and giant (dot-dashed) 
stars.}
\label{fig3}
\end{figure}

The presence of structures on the stellar radial density distribution (RDP, defined as 
$\sigma(R) = \frac{dN}{2\pi\,RdR}$) - or star-count density - may provide important clues on 
the dynamical state of a GC. M\,15, in particular, has evidence of being a post-core collapse
GC (\citealt{MarchiParesce96}) and, thus, the RDP of the very central region is critical. Here, 
we build the RDP with \rSDSS\ photometry, because this is the catalogue containing the largest 
number of stars with errors $<0.2$\,mag (Table~ \ref{tab1}). The J-PLUS RDP - based on the 
isophotal centre coordinates (J2000) $\alpha = 21^h:29^m:57^s.26$ and 
$\delta = 12\degr:09\arcmin:43\arcsec.92$ - covers the region from $\sim3\arcsec$ to well beyond 
the tidal radius (Fig.~\ref{fig3}).

As expected from ground-based and relatively low-pixel resolution observations, completeness 
effects (crowding and limited photometric depth) around the central part of the GC should prevent 
a large fraction of the stars from being detected, especially for $r\la3\arcmin$. This, in turn, 
would result in a stellar RDP presenting significant departures from classical profiles that 
usually describe well the large-scale RDP of GCs (e.g. \citealt{Miocchi2013}), such as the 
three-parameter \citet{King1962} and \citet{Wilson1975} profiles. These effects are clearly present 
in the J-PLUS RDP of M\,15 (Fig.~\ref{fig3}). However, we remark that most of the missing stars 
are MS ones, and the main focus hereafter will be the brighter sequences, like the RGB, AGB and 
HB (Sect.~\ref{JPLUS_CMD}).

Clearly, J-PLUS photometry is not adequate to probe the innermost region of such a populous and
crowded GC as M\,15 with the required resolution. Thus, the derivation of robust structural 
parameters of M\,15 requires to complement the J-PLUS data with higher spatial resolution ones. 
In this context, the high spatial resolution of HST would be ideal to build a complementary 
data-set to probe the central region of M\,15. An adequate data-set is the archival HST/ACS 
photometry of \citet{Sarajedini2007} and \citet{Anderson2008}, which contains stars brighter than
about 3 mags below the MSTO ($F814W\la22$). Given that this data-set fully encompasses the region 
$R\le0\arcmin.3$, the resulting RDP was directly (no scaling applied) transposed to Fig.~\ref{fig3}. 
Next, we used the archival HST/WFC3 data-set of \citet{Larsen2015} to sample the annulus within 
$0\arcmin.3\le R\le2\arcmin.3$ (this dataset excludes the central $R\la10\arcsec$ region of M\,15), 
which contains stars brighter than the sub-giant branch ($F555W\la19$). However, since this data-set 
provides only a partial coverage of M\,15's projected area, its RDP had to be multiplied by the factor 
7.1 to provide a perfect match with the ACS RDP. The combined ACS$+$WFC3 RDP merges smoothly into the 
J-PLUS one at $R\approx2\arcmin$. The composite HST$+$J-PLUS RDP (Fig.~\ref{fig3}) ends up sampling 
an unprecedented large radial range ($0\arcmin.12\le R\le50\arcmin$) with a relatively high quality.

\begin{figure}
\resizebox{\hsize}{!}{\includegraphics{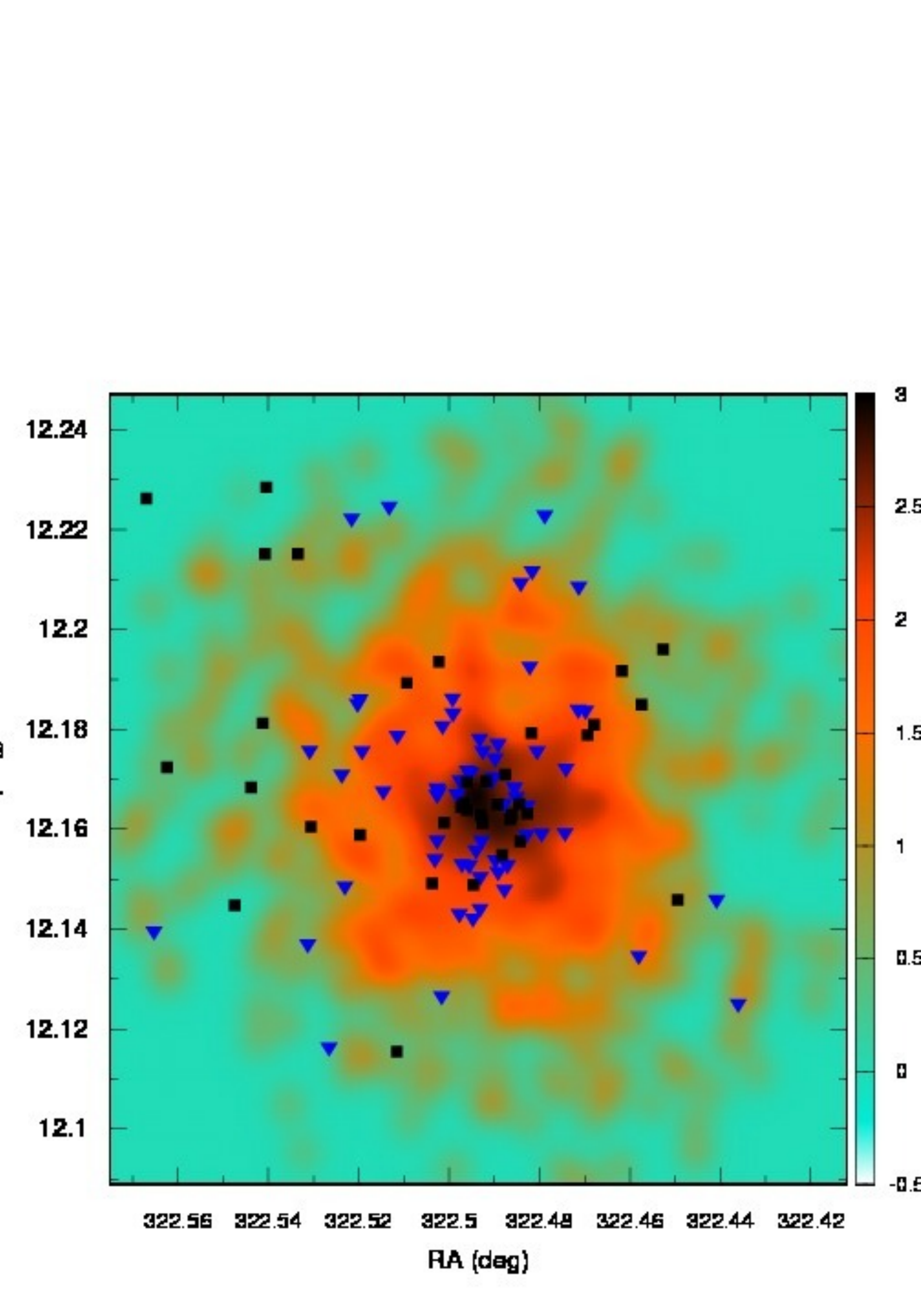}}
\caption[M\,15 colour maps]{Spatial distribution of the colour (\uJAVA\ - \zSDSS) 
across the inner $5\arcmin$ of M\,15. A redder colour near the centre is consistent with the 
relative central concentration of RGB and AGB stars. Stars distributed along the blue and red
sequences in Fig.~\ref{fig8} are shown as blue triangles and black squares, respectively.
Vertical bar shows the colour scale.}
\label{fig4}
\end{figure} 

Next, we fitted the classical three-parameter King profile\footnote{Defined as $\sigma(R)=\sigma_{bg} + 
\sigma_0\left[\frac{1}{\sqrt{1+(R/\rC)^2}} - \frac{1}{\sqrt{1+(\rT/\rC)^2)}}\right]^2$, where 
$\sigma_{bg}$ is the stellar-density background level.} to the composite RDP of M\,15 (excluding 
the completeness-affected portion of the J-PLUS RDP) as a simplifying assumption. The free parameters 
are the central density of stars ($\sigma_O$) and the background (sky) level, as well as the core 
(\rC) and tidal (\rT) radii. We obtain $\sigma_O\sim25~{\rm stars}\,arcsec^{-2}$ ($\sim1\times10^4$ 
stars ${\rm pc}^{-2}$), $\sigma_{bg}\approx7.2\times10^{-4}~{\rm stars}\,arcsec^{-2}$, 
$\rC=0\arcmin.15\pm0\arcmin.01$ ($\sim0.4$\,pc), and $\rT=28\arcmin.6\pm0\arcmin.9$ ($\sim83$\,pc). 
Integration of the RDP from the centre to \rT\ gives a number of present-day member stars of 
$\approx5.4\times10^4$. The values of \rC\ and \rT\ are consistent with previous determinations 
for M\,15 (e.g. \citealt{Harris2010}). 

To further probe RDP properties, now we consider samples of stars belonging to different luminosity 
(or equivalently, mass) classes: main-sequence and giants. Based, for instance on the CMD shown in 
Fig.~\ref{fig6}, we built samples of stars fainter and brighter than $g=18.2$. A similar criterion 
was applied to the HST data, in the sense that the stars corresponding to a given RDP have a 
similar luminosity range - although coming from different samples. Next we built the RDPs (according 
to the recipe given above for 
joining RDPs of different data sets) followed by giants and main-sequence stars (inset of Fig.~\ref{fig3}. 
It's clear that the spatial distribution of giants is more concentrated than that of the main-sequence 
stars, which is consistent with expectations of long-term mass-segregation effects acting on a GC (e.g. 
\citealt{McLFall08}; \citealt{Alessandrini16}). However, as a caveat we remark that number counts
for faint stars may suffer from radially variable incompleteness - especially at the central part - and, 
consequently, the amplitude of the difference between the two profiles can be overestimated. 

In Fig.~\ref{fig4} we show a colour (\uJAVA\ - \zSDSS) map of the high-stellar density region 
($R<5\arcmin$) of M\,15. Clearly, the central $\sim1\arcmin-2\arcmin$ is populated predominantly 
by stars with redder colours when compared to more external regions, which is consistent with the 
differences in density between the RDP built with all stars and that restricted to RGB$+$AGB stars 
(Fig.~\ref{fig3}).

\subsection{Surface brightness profiles}
\label{s_SBP}

\begin{figure}
\resizebox{\hsize}{!}{\includegraphics{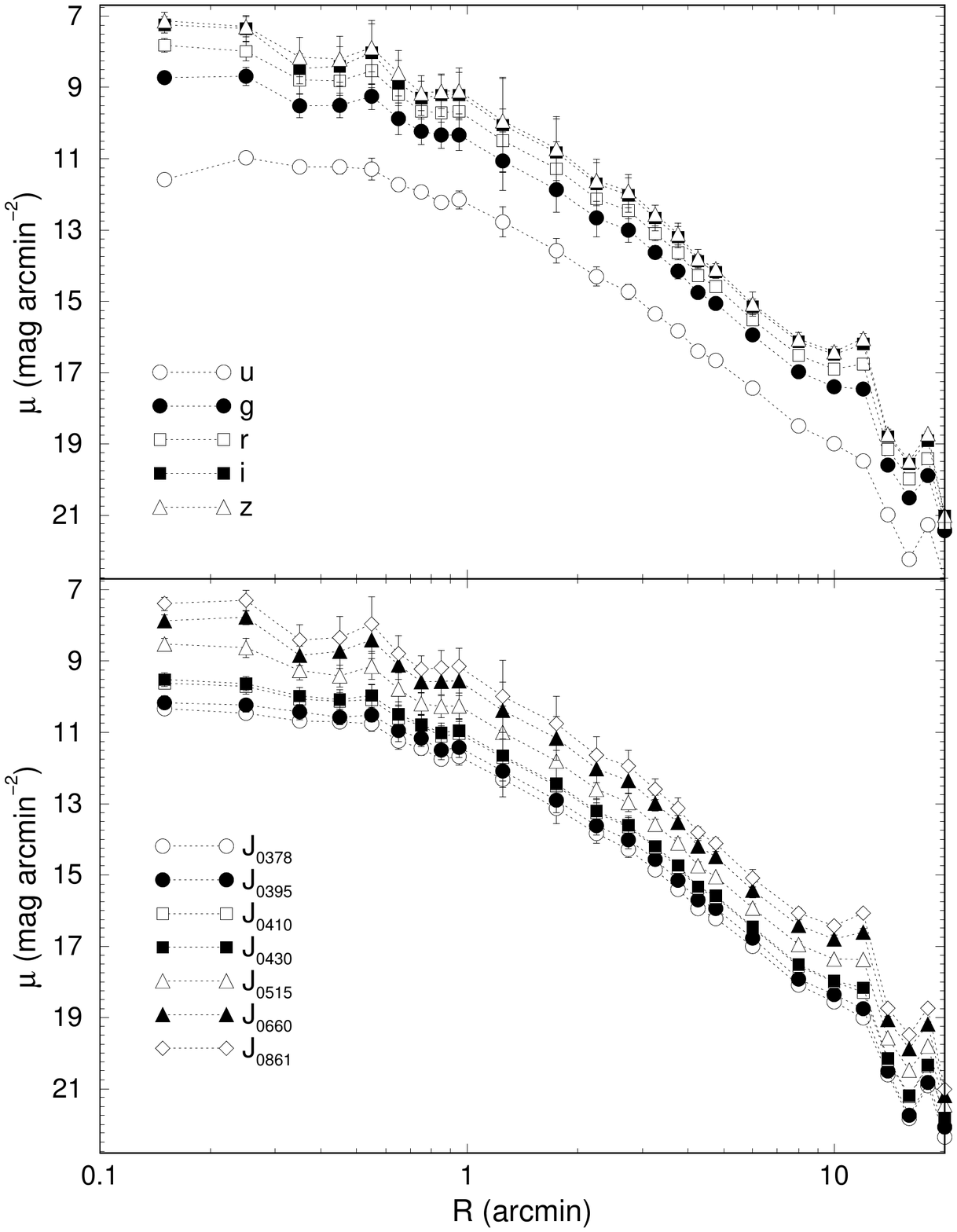}}
\caption[]{Surface brightness profiles for the narrow (top panel) and broad-band (bottom) 
filters.}
\label{fig5}
\end{figure}

Next we built the SBPs as a function of the distance to the 
cluster centre for all the filters (Fig.~\ref{fig5}). In general terms, SBPs are considered
as an easily obtained tool that can be used to derive the internal GC mass distribution by 
means of de-projection. Dynamical models of GCs usually require parameters such as the central 
surface brightness and the half-light radius, which can be derived from \citet{King1962} profile 
fits to the observed SBP. In the present case, the SBPs are built similarly to the RDPs 
(Sect.~\ref{s_RDP}), in the sense that the individual luminosities of all stars within concentric 
rings are summed up and later converted to surface brightness. Consequently, the final SBPs end up 
having a similar spatial resolution as the J-PLUS RDPs. The J-PLUS SBPs of M\,15 (Fig.~\ref{fig5}) are smoother and less sensitive 
to completeness than the star-counts corresponding to the RDP (Fig.~\ref{fig3}). 

\begin{table}
\caption{King fit to the J-PLUS SBPs of M\,15}
\label{tab2}
\renewcommand{\tabcolsep}{3.5mm}
\begin{tabular}{lccc}
\hline
\hline
Filter & $\mu_O$ & \rC  & \rT \\
       & ($mag\,arcmin^{-2}$) & (arcmin) & (arcmin) \\
\hline
\uJAVA    & $11.09\pm0.49$  & $0.38\pm0.10$   & $25.1\pm1.4$   \\
\Ja       & $9.47\pm1.08$   & $0.22\pm0.12$   & $25.0\pm1.3$   \\
\Jb       & $9.18\pm1.68$   & $0.21\pm0.17$   & $25.8\pm1.5$   \\  
\Jc       & $8.76\pm1.67$   & $0.20\pm0.16$   & $25.5\pm1.5$   \\  
\Jd       & $8.48\pm2.57$   & $0.17\pm0.21$   & $25.7\pm1.6$   \\ 
\gSDSS    & $7.91\pm1.44$   & $0.18\pm0.13$   & $25.2\pm1.0$   \\  
\Je       & $7.57\pm2.89$   & $0.16\pm0.21$   & $25.1\pm1.0$   \\   
\rSDSS    & $6.44\pm3.82$   & $0.11\pm0.21$   & $24.6\pm1.1$   \\   
\Jf       & $6.91\pm1.43$   & $0.16\pm0.11$   & $24.0\pm0.9$   \\ 
\iSDSS    & $5.96\pm3.08$   & $0.11\pm0.17$   & $24.8\pm0.9$   \\  
\Jg       & $5.73\pm1.08$   & $0.09\pm0.19$   & $24.6\pm1.2$   \\  
\zSDSS    & $6.16\pm0.49$   & $0.38\pm0.10$   & $25.1\pm1.4$   \\  
\hline
\end{tabular}
\begin{list}{Table Notes.}
\item $\mu_O$ is the central surface brightness.
\end{list}
\end{table}

Similarly to the RDP, the relatively coarse resolution of the telescope$+$camera system 
(Sect.~\ref{Data}) does not allow the J-PLUS SBPs to resolve in detail the collapsed core 
of such a distant GC as M\,15. Nevertheless, they can be adequately fitted by the 3-par 
King profile. The resulting parameters are in Table~\ref{tab2}. Despite the significant 
uncertainties - especially in \rT - the structural radii derived from the SBPs are very similar
to the HST$+$J-PLUS composite RDP ones (Fig.~\ref{fig3}).

\section{J-PLUS colour-magnitude diagrams}
\label{JPLUS_CMD}

Among other issues, the presence of multiple populations in GCs has raised great 
interest in the recent literature (e.g. \citealt{Gratton2012}). The wide-field T80 photometry, 
complemented with its special filter system, may be a valuable tool to reveal and characterize 
MPs in colour-magnitude diagrams (CMDs).

\begin{figure}
\resizebox{\hsize}{!}{\includegraphics{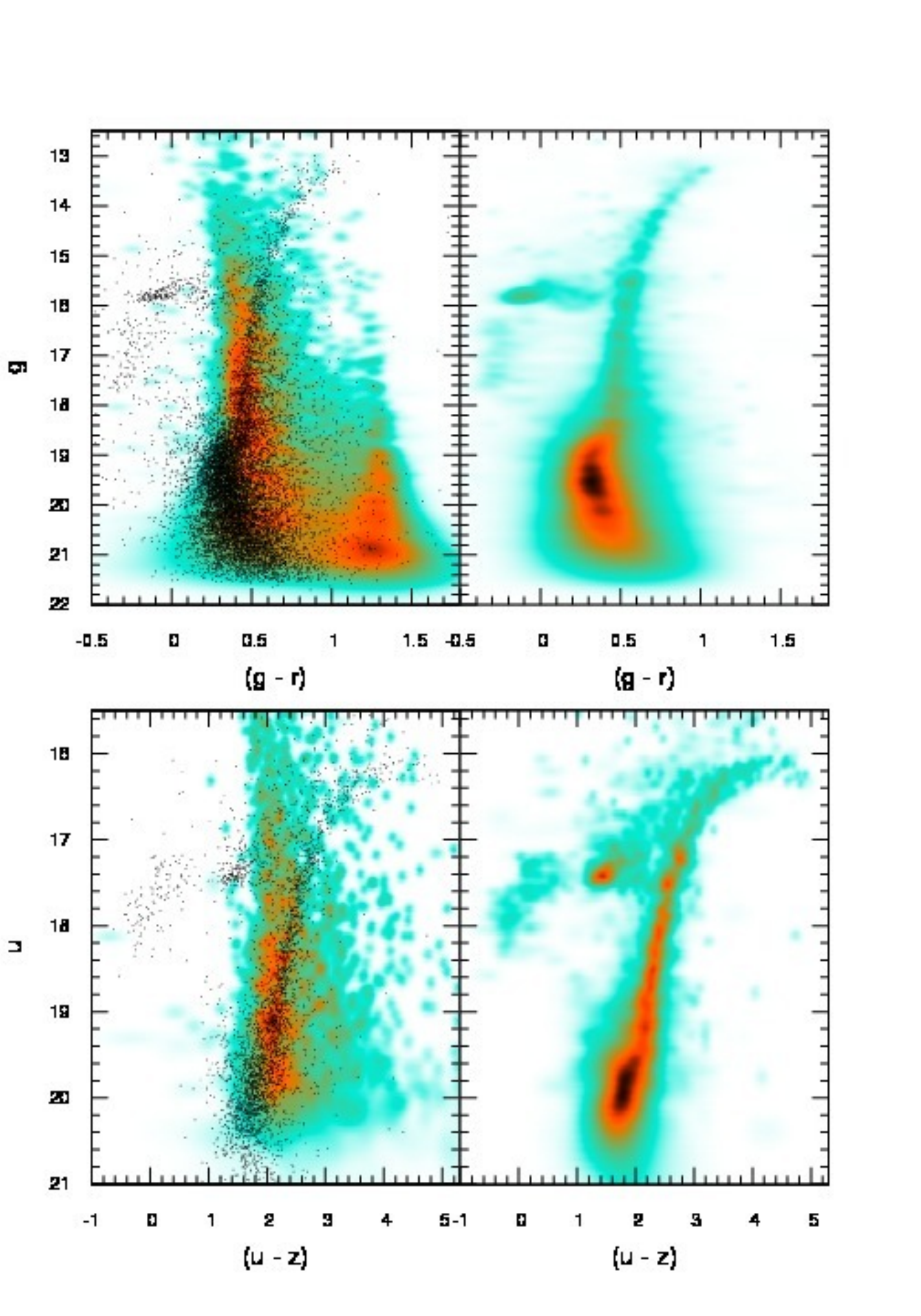}}
\caption[]{Field-star decontamination illustration in different colours. Left panels: The Hess 
diagram of the underlying field stars is compared to the CMDs extracted in the inner 5\arcmin\ of 
M\,15 (small black points). Right: Hess diagrams built with the decontaminated photometry.}
\label{fig6}
\end{figure}

\begin{figure}
\resizebox{\hsize}{!}{\includegraphics{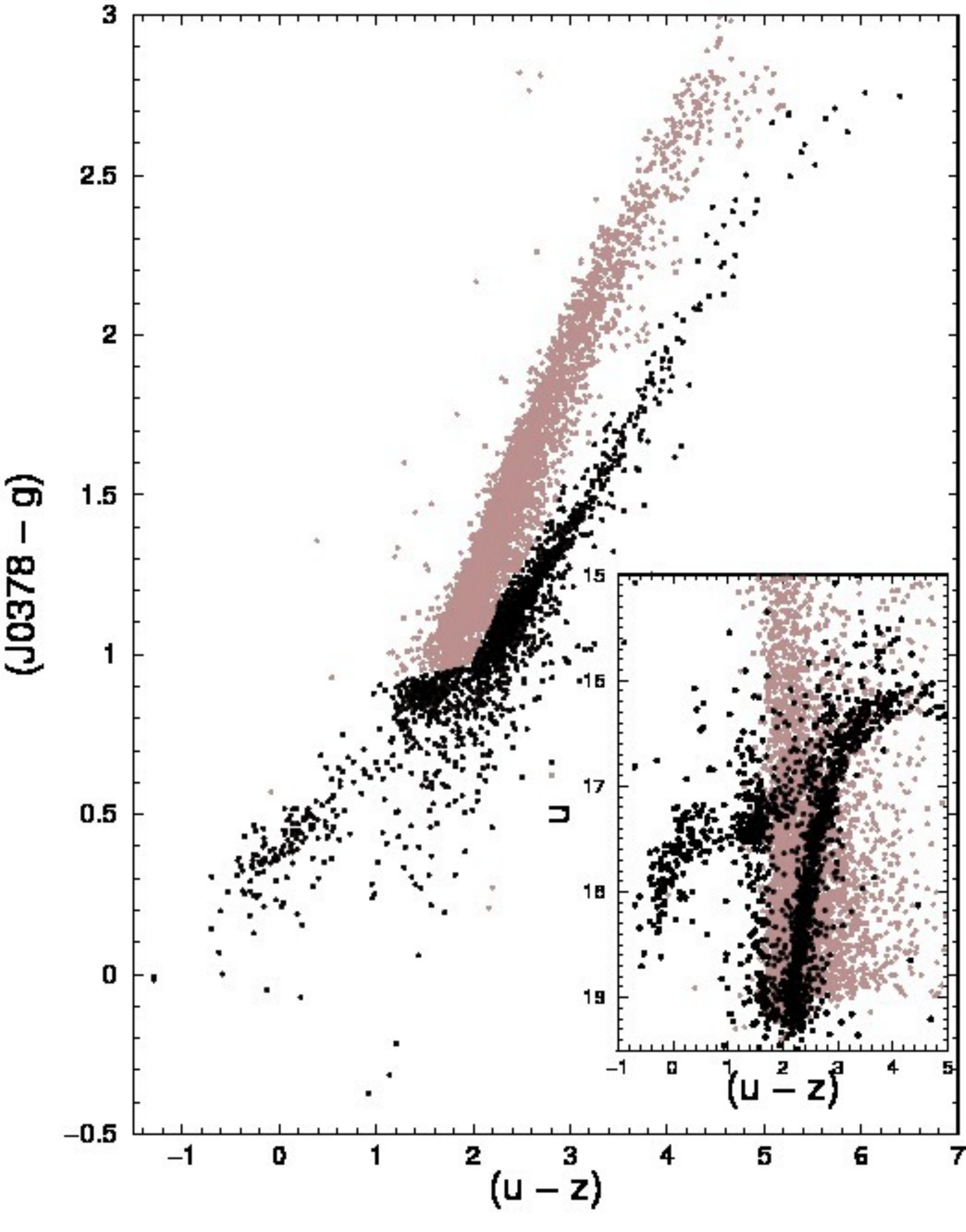}}
\caption[]{(\uJAVA - \zSDSS) vs. (\Ja - \gSDSS) colour-colour diagram for the stars located 
within 5\arcmin\ from the centre of M\,15. Stars in each sequence correspond to M15 (black 
points) and field (brown). Their locations in the corresponding CMD are shown in the inset.} 
\label{fig7}
\end{figure} 

Despite its location at a relatively high Galactic latitude ($b=-27.31\degr$), a significant
number of field stars is present across the area on the sky occupied by M\,15, and the wide 
FoV of T80Cam - including a wide area beyond its tidal radius that can be used as comparison 
field - is adequate to take their contamination into account. As already implied by the RDP 
(Fig.~\ref{fig3}), the contamination is not expected to be excessive, at least at the central 
part of M\,15. In any case, we applied the decontamination algorithm developed by \citet{Bonatto07} 
to minimize the presence of field stars on CMDs and produce cleaner photometry. In summary, the 
algorithm assumes that the colour and magnitude distribution of the contaminant stars across the 
cluster area is the same as that displayed by the offset field, with differences essentially related 
to the relative densities. Thus, the algorithm divides the observed colour and magnitude space (of 
the cluster and the area-corrected offset field) in a grid of cells and builds the respective Hess 
diagrams. Then it computes the cluster intrinsic stellar density, which is simply the difference 
between both Hess diagrams. The cell size of the Hess diagrams are considerably wider than the 
photometric errors but small enough to preserve the morphology of the evolutionary sequences. 
The efficiency of this approach relies heavily on the colour and magnitude representativity of 
the offset field with respect to the contaminant stars. In Fig.~\ref{fig6} we illustrate the decontamination 
process for two different colours; one exclusive to J-PLUS (\uJAVA - \zSDSS) and the widely used 
(\gSDSS - \rSDSS). Field contamination accounts only for $\sim2\%$ and $\sim9\%$ of the stars 
present, respectively in the CMDs \gSDSS $\times$ (\gSDSS\ - \rSDSS) and \uJAVA $\times$ 
(\uJAVA\ - \zSDSS) in the central 5\arcmin\ region of M\,15, which is consistent with the high 
density contrast between the RDP in this region and that beyond the tidal radius (Fig.~\ref{fig3}). 
These fractions increase to $\sim58\%$ and $\sim40\%$ in the region 10\arcmin-20\arcmin. Field
stars in both CMDs have colours and magnitudes distributed mainly across the RGB (and main-sequence 
turnoff - MSTO) of M\,15. Some contamination  - especially for bright stars - remain in the final
CMDs.

We note that, instead of the discrete CMDs, in what follows we will mostly use their continuous
counterpart, the Hess diagrams. Hess diagrams \citep{Hess24} contain the relative density of 
occurrence of stars in different colour-magnitude cells of the Hertzsprung-Russell diagram; 
photometric uncertainties are explicitly taken into account in the density computation. In short, 
the CMD is divided into a grid of cells, and the fraction of the colour and magnitude of all stars 
is computed for each cell. Compared to CMDs, Hess diagrams provide a clearer view of the evolutionary 
sequences, especially in those having a large number of (usually superimposed) stars. The 
decontaminated Hess diagrams are typical of a very metal-poor GC, displaying a prominent blue-HB 
and reaching the upper RGB. The main-sequence turnoff is detected in both filters, but especially 
the \gSDSS.

Interestingly, among the several possible colour-colour combinations, we found one that separates 
field from M\,15 stars. In Fig.~\ref{fig7} we show the colour-colour diagram (\uJAVA - \zSDSS) vs. 
(\Ja - \gSDSS) for all stars present in the inner 5\arcmin. We note that there are two visible sequences 
for colours redder than $\sim 1$, but merging together for bluer values in both colours. As shown in 
the corresponding CMD (inset), each sequence corresponds to either of the expected stars of M\,15 and 
the field. We note that the separation occurs more clearly for stars brighter than the bottom of the RGB; 
field and M\,15 main-sequence stars merge together. If the separation between these two sequences holds 
for GCs of any metallicity and/or projected towards any direction in the Galaxy, such a colour-colour 
diagram may become a valuable discriminator of field and GC stars. The GC and field stars colour-colour 
separation will be subsequently verified for GCs with higher metallicity and different directions than 
M\,15. However, we remark that, contrarily to the quantitative decontamination algorithm applied above
(Fig.~\ref{fig6}), the colour-colour separation of sequences does not take the relative stellar 
density (GC member to field stars) into account. In this sense, it serves only to provide a qualitative 
perspective of the intrinsic morphology of the evolutionary sequences of a given GC. In the case of 
M\,15, it separates only stars brighter than the main-sequence turnoff. 

\subsection{Can we detect multiple populations?}
\label{multipop}

Spectroscopically, MPs are characterized by populations of stars having different chemical compositions, 
particularly of the light elements, including He, C, N, O, and Na (e.g. \citealt{Milone2014},
\citealt{Gratton2012}, \citealt{Pancino2010}). Moreover, depending on the photometric quality, MPs can 
show up as splits in the evolutionary sequences from the main sequence to the giant branch (e.g.
\citealt{Gratton2012}). Quite recently, \citet{Gruyters2017} report the first ever detection of MPs along 
the AGB of NGC\,6752, obtained with Str\"omgren photometry. 

Coupling photometry from HST/WFC3 (for the centre) and SDSS (for $R\ga1$\,pc) to chemical enrichment
models, \citet{Larsen2015} showed evidence of multiple populations (characterized by N-normal 
and N-enhanced compositions) of red giants that are differently distributed in M\,15, at least in the 
radial range $4\arcsec - 130\arcsec$. Exploiting the well-established sensitivity 
of UV photometry to light-element abundance variations - accessible with WFC3/HST filters - they 
found that giants in the lower RGB with primordial composition are more centrally concentrated, 
with a reversal of the trend for radii greater than 1\arcmin. \citet{Larsen2015} also found that 
the observed spread of the RGB stars in the colour ($F_{343N} - F_{555N}$) is significantly larger 
than that expected from photometric uncertainties, thus likely characterizing intrinsically different 
populations of stars.

\begin{figure}
\begin{minipage}[b]{0.5\linewidth}
\includegraphics[width=\textwidth]{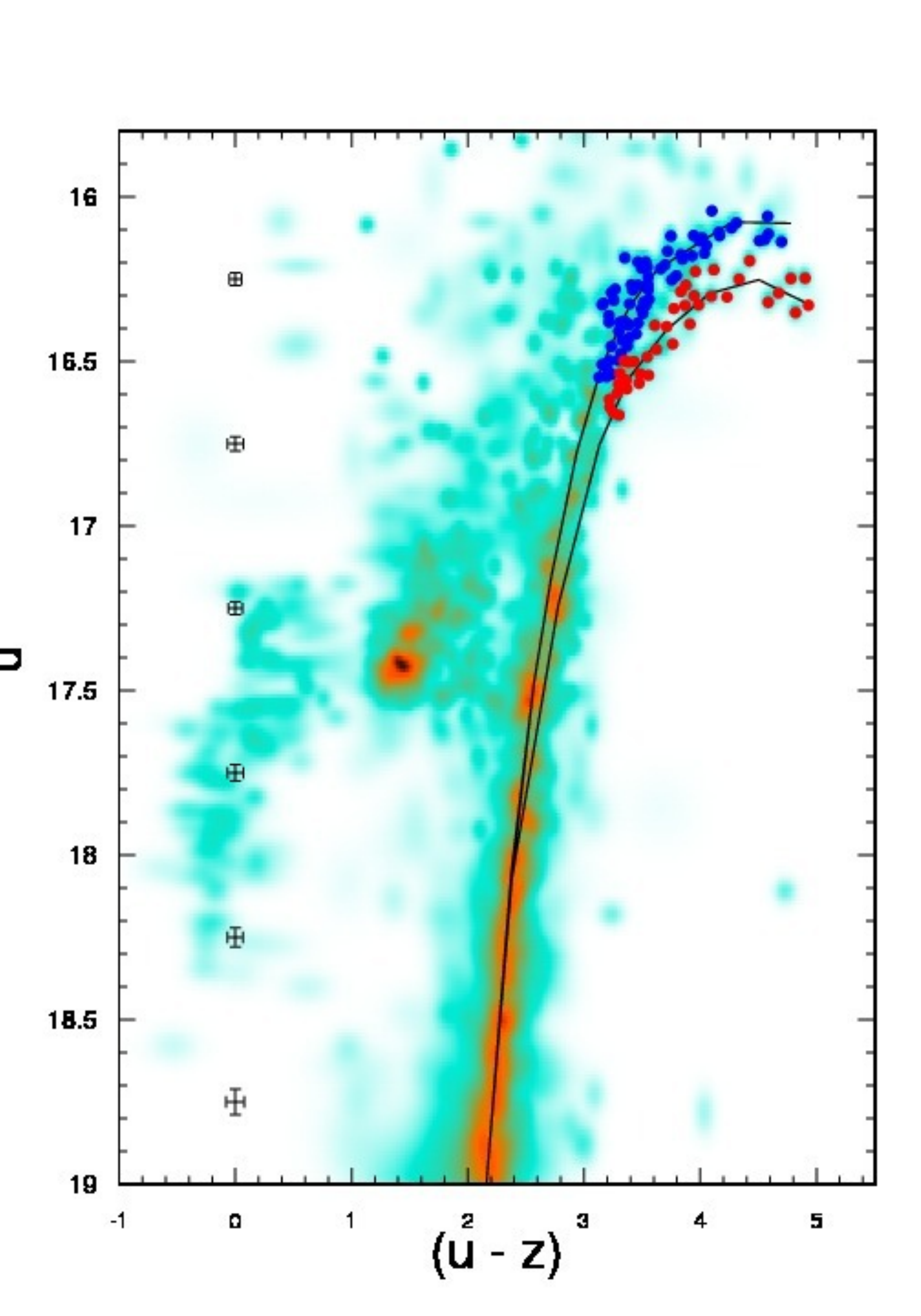}
\end{minipage}\hfill
\begin{minipage}[b]{0.5\linewidth}
\includegraphics[width=\textwidth]{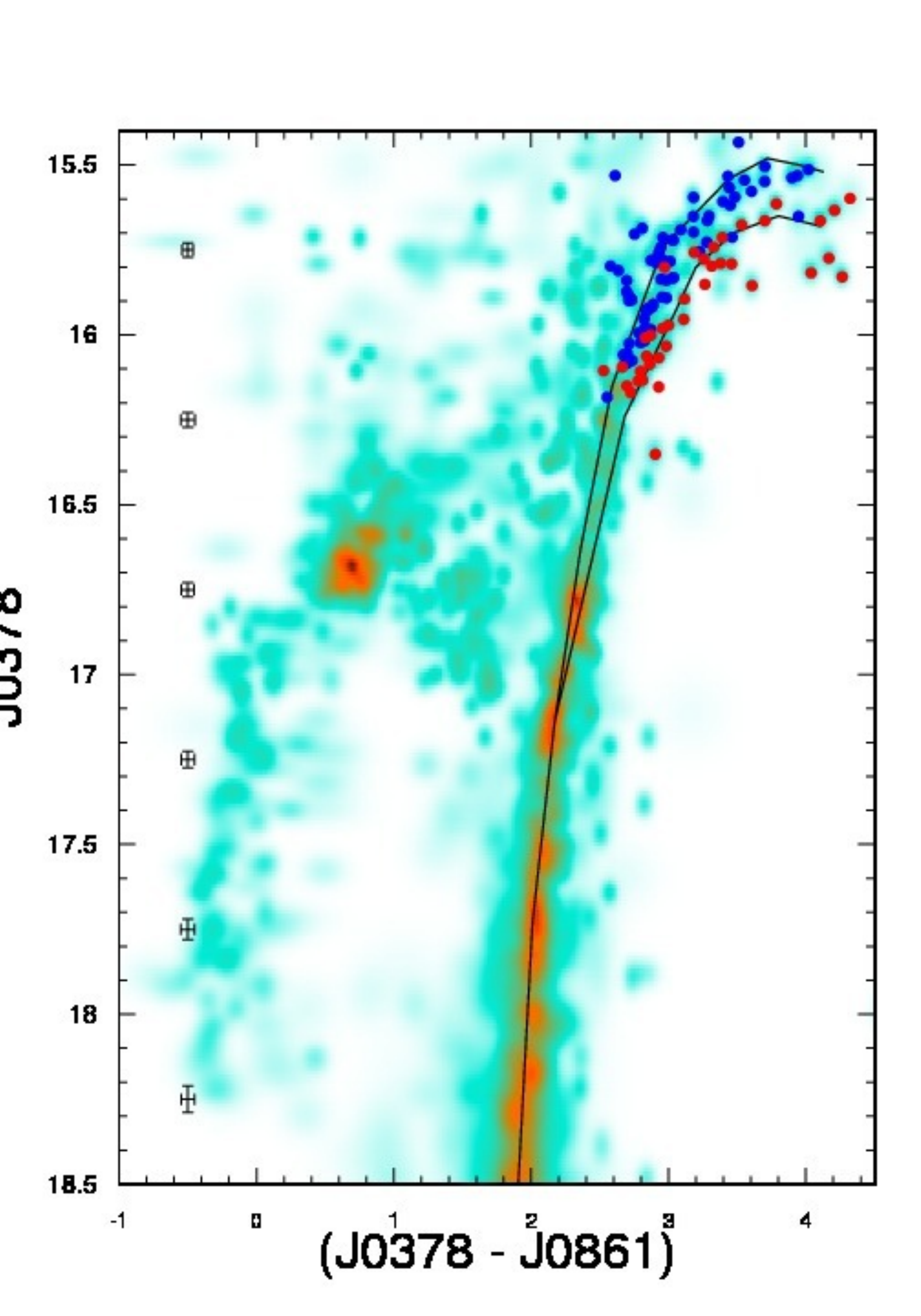}
\end{minipage}\hfill
\begin{minipage}[b]{0.5\linewidth}
\includegraphics[width=\textwidth]{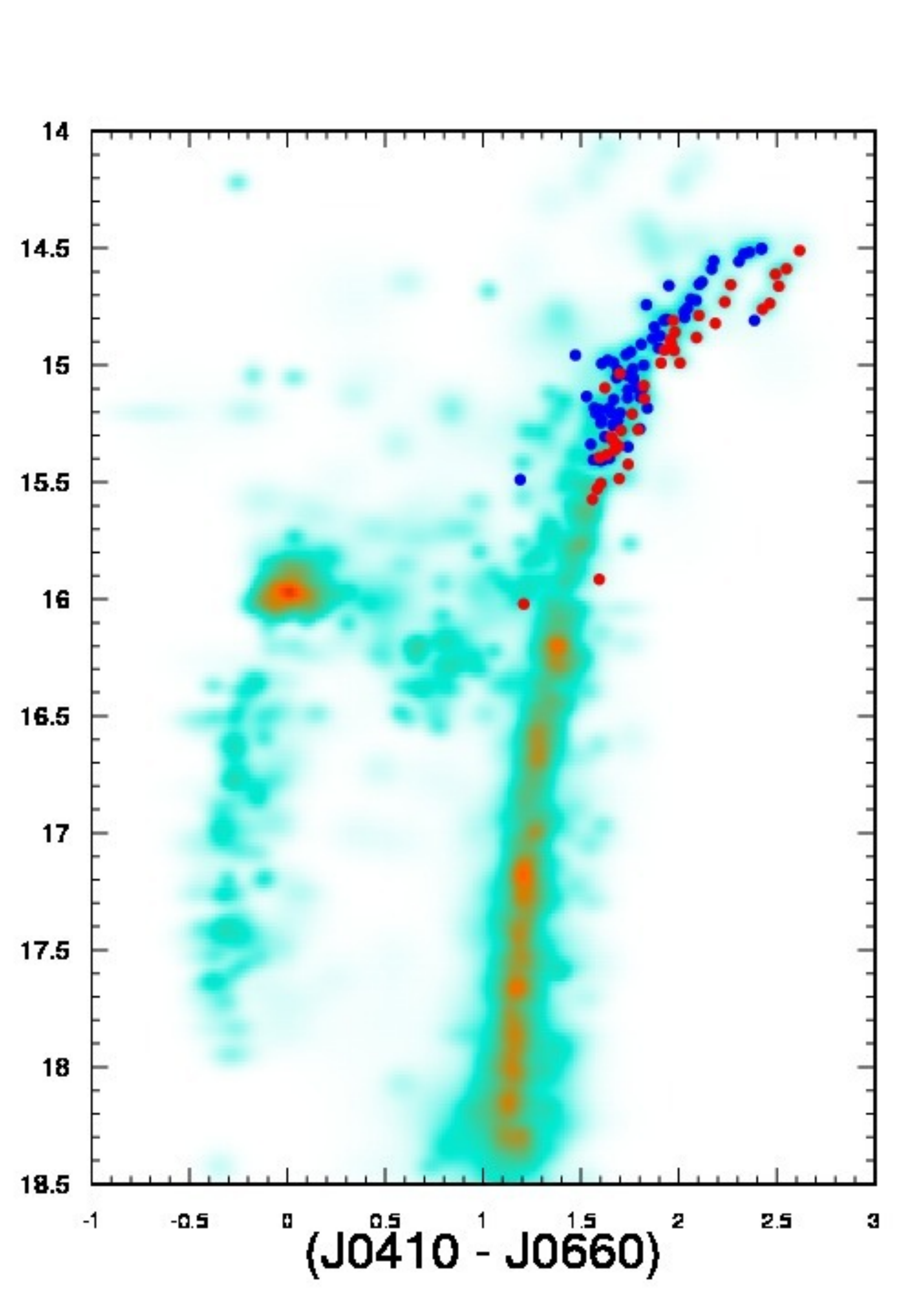}
\end{minipage}\hfill
\begin{minipage}[b]{0.5\linewidth}
\includegraphics[width=\textwidth]{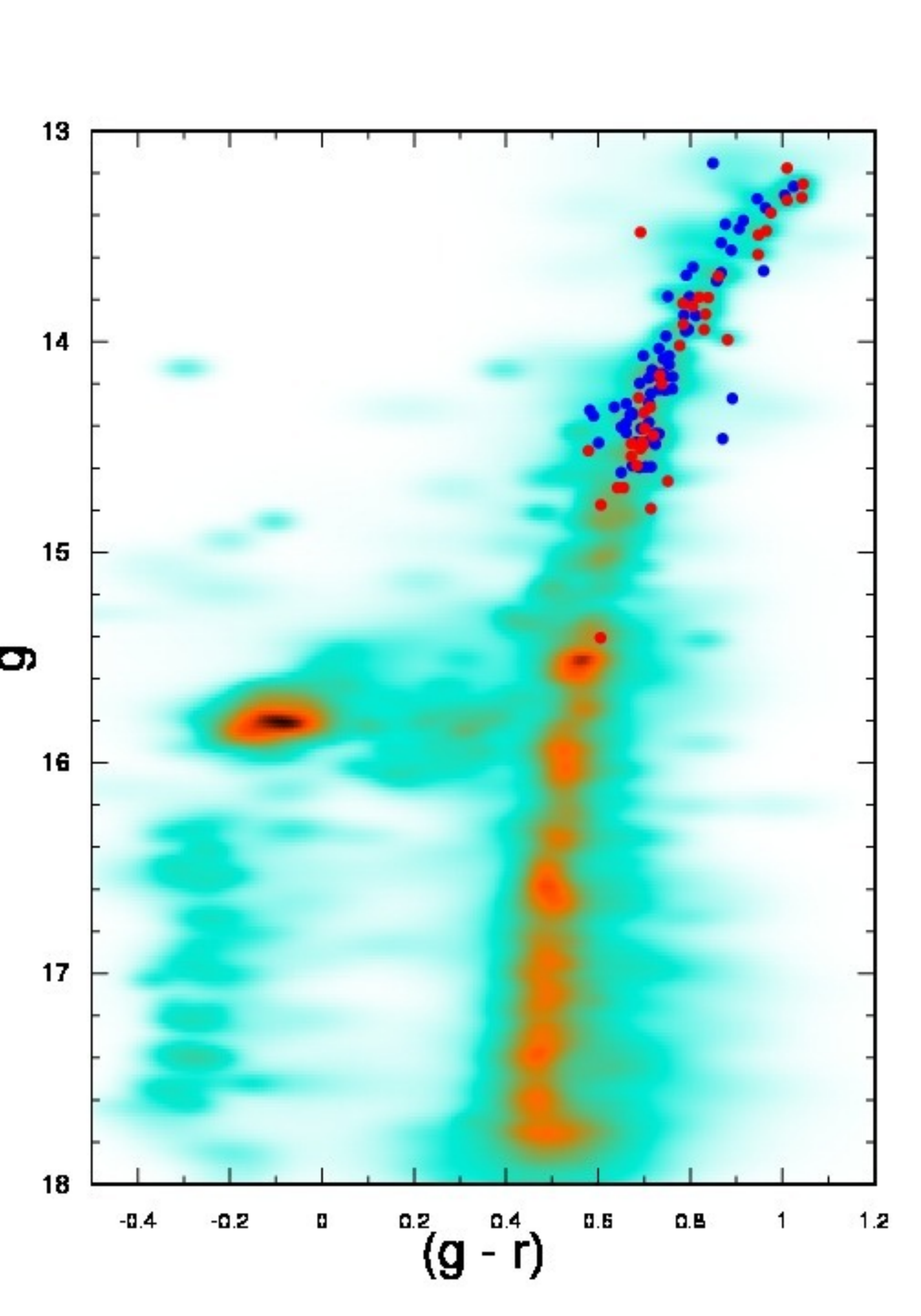}
\end{minipage}\hfill
\caption[]{Top panels: The CMDs that best show evidence of more than one population, extracted 
from the central 5\arcmin\ of the cluster. Fiducial lines for the apparent sequence split in the 
upper RGB are shown, together with the average error bars computed in bins 0.5\,mag wide. 
Stars of the blue and red sequences are overplotted on the Hess diagram. Bottom: same as above for
combinations of filters that are not much sensitive to light element abundance variations.}
\label{fig8}
\end{figure}

The spectral distribution of the 12 J-PLUS bands may provide an adequate means to distinguish 
between atmospheric effects related to light-element abundance variations and those impacting the 
stellar structure. In this context, we have used the J-PLUS bands that are more sensitive to metallic 
bands (Table~\ref{tab1} and Fig.~\ref{fig11}) to build CMDs reaching to the upper RGB and AGB 
(Fig.~\ref{fig6}), to search for the colour that best presents evidence of sequence splits. This 
separation is more clearly visible with the bluest and reddest J-PLUS filters, $\uJAVA\times (\uJAVA - \zSDSS)$ 
followed by $\Ja\times (\Ja - \Jg)$. In the top panels of Fig.~\ref{fig8} we show the Hess diagrams 
corresponding to the CMDs (\uJAVA\ vs. (\uJAVA - \zSDSS) and \Ja\ vs. (\Ja - \Jg) extracted from the 
central 5\arcmin, with field-decontaminated photometry. We note that the RGB appears to split into two 
sequences from $\uJAVA\la17$ and $\Ja\la16.5$. The splits are considerably wider than what would be 
expected from photometric scattering, as can be seen by the average error bars (Fig.~\ref{fig8}). 
Fiducial lines - arbitrarily built to represent the mean path followed by each sequence - are included 
in Fig.~\ref{fig8} as a visual aid.

Visually considering the sequence split and guided by the fiducial lines in the $\uJAVA\times 
(\uJAVA - \zSDSS)$ CMD, we separated the stars in blue and red sub-samples, and took their magnitudes 
in the other J-PLUS filters. We remark that when these stars are directly transposed into the $\Ja\times 
(\Ja - \Jg)$ CMD, most of them consistently fall in the corresponding (blue and red) sequences. 

To check whether any potential systematic effect might have been introduced on the images 
by the 2D-calibration process, thus producing gradients or affecting one corner more than the others,
we plot in Fig.~\ref{fig4} the locii occupied by the upper RGB$+$AGB stars that define the blue and 
red sequences seen in Fig.~\ref{fig8}. It turns out that both samples are more or less equally randomly 
distributed across the image, thus ruling out systematic calibration effects as a possible cause of 
the split. 

\begin{figure}
\resizebox{\hsize}{!}{\includegraphics{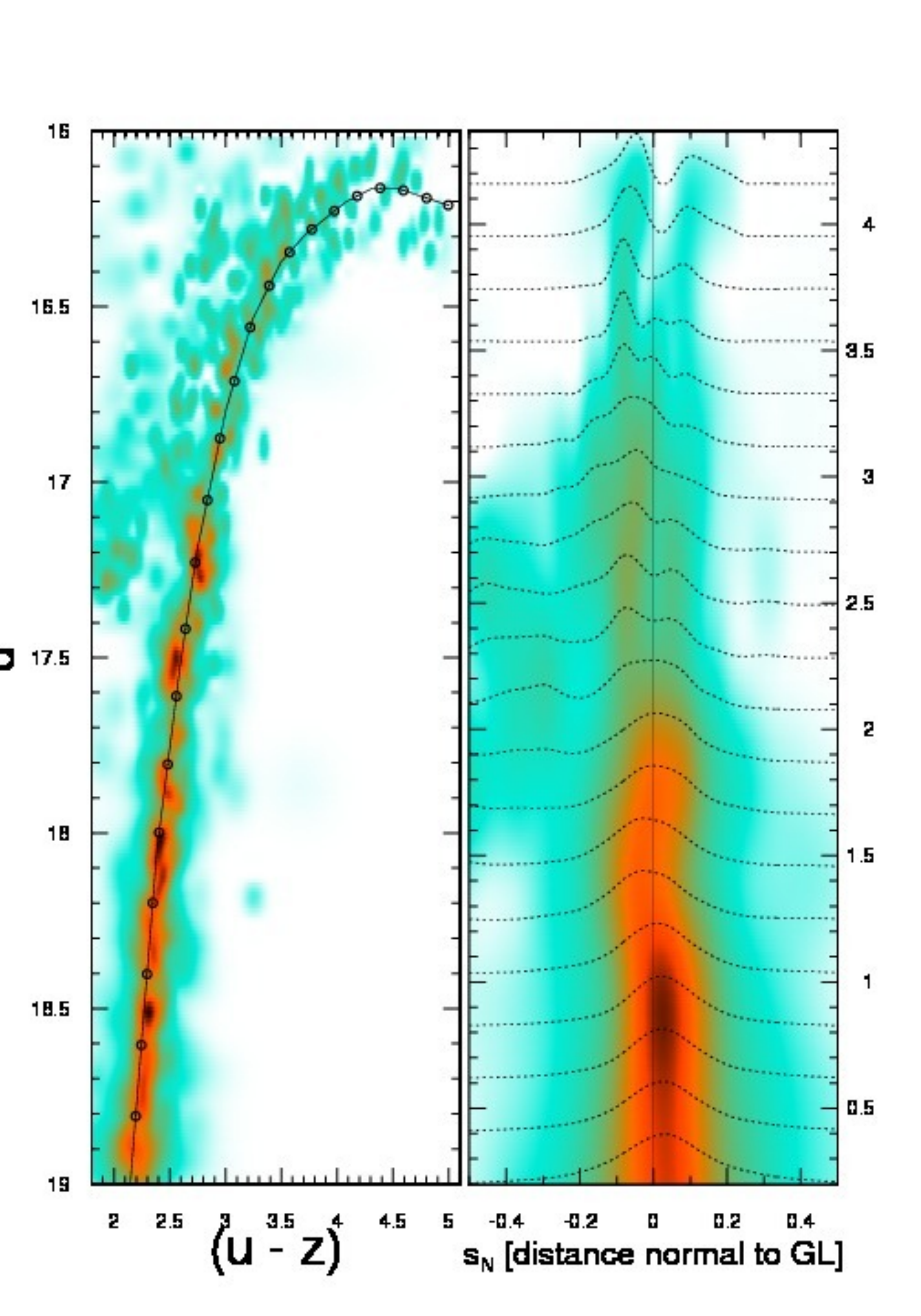}}
\caption[]{Left panel: portion of the CMD to be rectified along the guiding (continuous) line (GL);
the position of the individual sectors along the GL is marked (circles). Right: rectified stripe 
($\approx1$ mag wide) of the RGB/AGB overplotted with the stellar density profiles (dotted lines) 
extracted from sectors perpendicular to the GL; distance along the GL is marked on the right axis; 
profiles are normalized to their peak value for visualization purposes. }
\label{fig9}
\end{figure}

The presence of the split along the RGB can be better visualized using rectified evolutionary 
sequences. We do this by first building a guiding line (GL), which is the median between the fiducial 
lines corresponding to the blue and red stellar sub-populations shown in the top-left panel of 
Fig.~\ref{fig8}. Then we divide the GL in a series of segments (of same size) and, for each segment 
we compute the stellar density in the direction perpendicular to the segment, in both sides of the GL. 
In practice, we compute the angle of the segment direction (in the CMD) with respect to the vertical, 
and apply the corresponding rotation to the position (colour and magnitude) of all stars (and uncertainty). 
The final result is shown in Fig.~\ref{fig9}, which also shows the portion of the CMD to be rectified 
and the stellar-density profiles (1\,mag wide) extracted from bins projected in the direction perpendicular 
to the GL. The split appears to start at a distance along the GL $\approx2.2$, which corresponds to 
$\uJAVA\sim17$. 

\begin{figure}
\resizebox{\hsize}{!}{\includegraphics{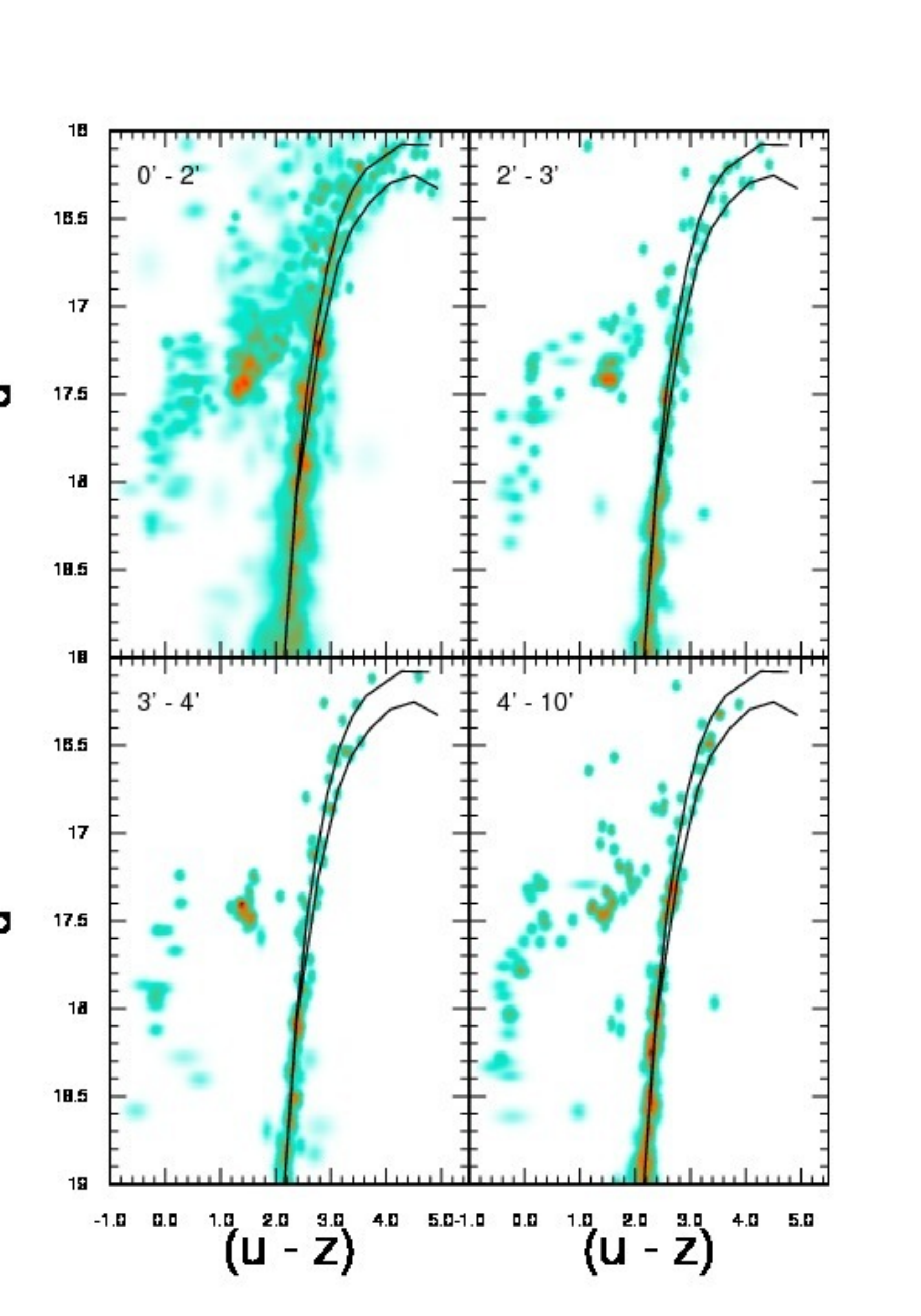}}
\caption[]{CMDs extracted from different regions to show that the RGB split extends up to $\sim4\arcmin$ 
from the cluster centre. }
\label{fig10}
\end{figure}

\begin{figure}
\resizebox{\hsize}{!}{\includegraphics{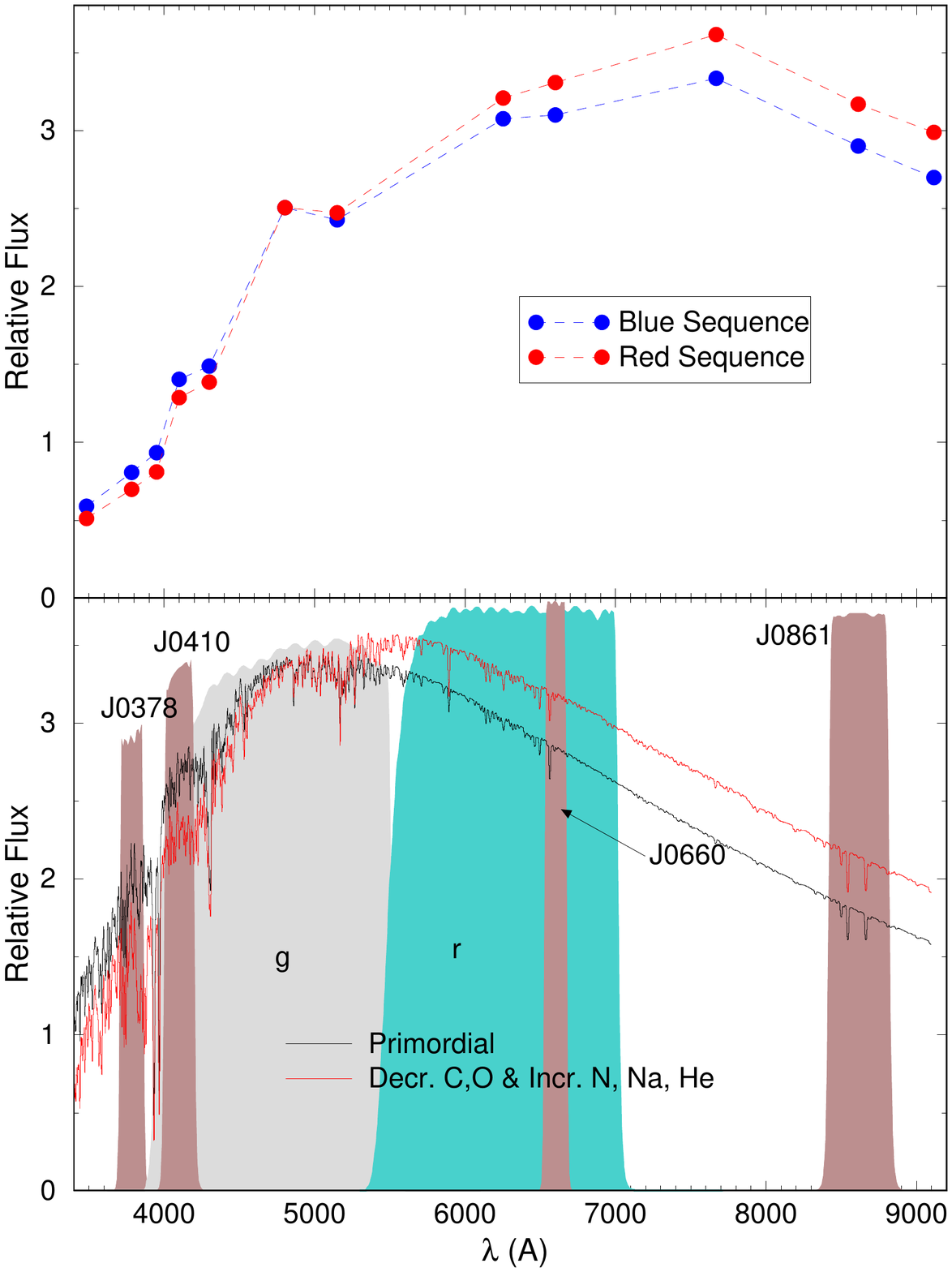}}
\caption[]{Top: Average J-PLUS SEDs for the stars in the blue and red sequence splits. Normalized at 
\gSDSS\ for comparison purposes. Bottom: Synthetic spectra by \citet{coelho+11} and \citet{coelho+14}
representing a RGB of a primordial sub-population in black (${\rm T_{eff}} = 4500$\,K, $\log g = 1.5$, 
$\feH = -2.3$ dex, [$\alpha/Fe] = 0.4$\,dex, $Y = 0.256$), and a slightly cooler giant representing a 
secondary population in red (${\rm T_{eff}}=4250$\,K, $\log g=1.5$, $\feH=-2.3$\,dex, [$\alpha/Fe] = 0.4$ 
dex, $[C/Fe] = -0.30$\,dex, $[N/Fe] =+1.20$\,dex, $[O/Fe]=-0.45$\,dex, $[Na/Fe]=+0.60$\,dex, $Y = 0.300$). 
J-PLUS filters used in Fig.~\ref{fig8} are highlighted.}
\label{fig11}
\end{figure}

Next, we investigate (Fig.~\ref{fig10}) the radial dependence of the upper-RGB split by means 
of CMDs extracted from different radial bins for $\uJAVA\times (\uJAVA - \zSDSS)$. The fiducial lines 
built for the full 5\arcmin\ region CMD are directly overplotted in the CMDs built for the  
$0\arcmin-2\arcmin$, $2\arcmin-3\arcmin$, $3\arcmin-4\arcmin$ and $4\arcmin-10\arcmin$ radial bins, 
with no shifts applied. One can conclude that the split can be detected at least up to 3\arcmin\ 
from the centre of the cluster, and maybe reaching 4\arcmin. This is expected, as the RGB stars are 
essentially concentrated towards the centre of the cluster (see Fig.~\ref{fig3}).

To further search for clues to the origin of the detected split, we build the average J-PLUS SEDs 
for the stars in each sequence. Magnitudes in each filter are transformed into flux, averaged out 
and normalized to the flux at \gSDSS\ for comparison purposes. The result is shown in Fig.~\ref{fig11}. 
On average, blue-sequence stars have higher flux than red-sequence ones for $\lambda\la 4500$\,\AA, 
with the trend consistently reversing otherwise. This is consistent with the blue sequence stars being 
systematically slightly hotter than the red sequence stars. This is the expected behaviour if we 
interpret the blue-sequence stars as being RGBs from the primordial sub-population of the cluster, 
while the red-sequence would correspond to the sub-population of RGBs with altered chemistry for C, 
N, O and Na - see, for example, Figs.~1 and 2 in \citealt{Larsen2015}, which show qualitatively the same 
behaviour, but for different - but analogous - bands. It is clear that the flux (and magnitude) 
differences between both sequences are systematic and not related to a particular band. In the lower 
panel of Fig.~\ref{fig11} we further explore qualitatively this interpretation. We show synthetic 
models of RGB stars computed as in \citet{coelho+11} and \citet{coelho+14}, degraded to medium 
resolution for easier visualization. The spectrum in black is from an RGB star with the chemistry 
of a primordial sub-population in a cluster with similar metallicity as M\,15 (${\rm T_{eff}}=4500$\,K, 
$\log g=1.5$, $\feH=-2.3$ dex, [$\alpha/Fe] = 0.4$\,dex). The red one is from a cooler RGB 
(${\rm T_{eff}}=4250$\,K, same $\log g$ and \feH\ as the black spectrum) but with variations in the 
abundances of C, N, O, Na and He. The similarity between the observed and the synthetic SEDs is seen 
clearly. Although just qualitatively, this analysis may provide some clues to what is causing the 
split. Additionally, we remark that stellar evolution models have shown that an increase in the 
abundance of He changes the overall properties of the stellar population by increasing ${\rm T_{eff}}$ 
of the stars, while C, N, O, and Na do not (e.g. \citealt{Cassisi2017}). This effect will in turn also 
affect spectroscopy sensitive to Hydrogen lines (e.g. \citealt{coelho+11}). Considering the above, we 
find evidence that pseudo-spectral fitting has potential to trace chemical variations among stars, though 
a quantitive analysis will be explored in future work.

Besides being sensitive to stellar temperature, some of the J-PLUS filters may also be sensitive
to metallicity. For instance, \Ja\ (and \uJAVA) overlaps the NH molecule in the near-UV, while \Jg\ 
(and \zSDSS) encompasses the Ca\,II Triplet in the near-IR. In this sense, the colours $(\Ja - \Jg)$ 
and $(\uJAVA - \zSDSS)$ may be sensitive to differences in the N and Ca abundances. The wavelength 
range covered by \uJAVA\, also includes features known to be sensitive to the variations of C and N 
typical of clusters with multi-populations \citep{coelho+11,coelho+12}. Exploring the sensitivity 
of the WFC3/HST filters F275W, F336W, and F438W to the molecular bands OH, NH, and CN $+$ CH, 
respectively, \citet{Piotto2015} have shown that the pseudo-colour 
$C=(m_{F275W} - m_{F336W}) - (m_{F336W} - m_{F438W})$ splits the RGB of M\,15 in at least 2 
sequences (see their Fig.~22). 

As another suggestion that the split observed in the $\uJAVA\times (\uJAVA - \zSDSS)$ and $\Ja\times 
(\Ja - \Jg)$ CMDs is likely related to light-elements abundance, in the bottom panels of
Fig.~\ref{fig8} we show
the CMDs $\Jc\times (\Jc - \Jf)$ and $\gSDSS\times (\gSDSS - \rSDSS)$. \Jc\ and 
\Jf\ overlap H$\delta$ and H$\alpha$, respectively, while both \gSDSS\ and \rSDSS\ are very broad 
and encompass several lines and continuum. There is still some separation (albeit marginal) between 
the blue and red samples in the $\Jc\times (\Jc - \Jf)$ CMD, but much less clear the in the previous 
CMDs. However, they become mostly mixed up in the $\gSDSS\times (\gSDSS - \rSDSS)$ CMD, thus suggesting 
that the separation is indeed related to the light-elements abundance variations.

In summary, it seems quite plausible that additional J-PLUS filter combinations - such as \Ja\ and \Jg - can separate
MPs in Galactic GCs, even for such a low-metallicity GC as M\,15. Thus, it would be interesting to 
investigate the effect on the J-PLUS filters produced by a GC hosting two sub-populations characterized
by the same overall metallicity but presenting topical abundance changes. Since M\,15 is very
metal-poor, the next obvious target is a metal-rich GC, and \citet{Harris2010} contains 22 GCs more 
metal-rich than $\feH=-1.0$ and closer than $d_\odot\sim10$\,kpc from the Sun.

\section{Concluding remarks}
\label{conclu}

In this paper we present science verification (or Phase\,III) data obtained with the 12 J-PLUS 
filters for the M\,15 globular cluster. This data set was obtained as a test-case to investigate 
the potential applications of J-PLUS to the study of Galactic GCs by means of the instrument's 
wide field capabilities and the full set of narrow and broad-band filters photometry. 

M\,15 is very metal-poor ($\feH=-2.3$) and relatively distant from the Sun ($d_\odot\sim10$\,kpc).
Even so, the filter combination $(\uJAVA - \zSDSS)$ is able to show the presence of MPs by
means of an upper-RGB split. $(\Ja - \Jg)$ also shoes similar behaviour, though witj lower
significance. In the case of M\,15, 
the split is detected at least up to $\sim3\arcmin$ ($\sim8.7$\,pc) from the cluster centre. 

A follow-up to the findings presented here is underway using similar observations of closer 
and less metal-poor GCs. In fact, \citet{Harris2010} catalogue lists 22 GCs more metal-rich than 
$\feH=-1.0$ and closer than $d_\odot\sim10$\,kpc from the Sun. All of them can be observed
by S-PLUS\footnote{The Southern Photometric Local Universe Survey (S-PLUS) is a joint scientific 
effort of Brazilian, Chilean and Spanish institutions to map $\sim8000\deg^2$ of the southern 
sky with the same filters as J-PLUS, expected to have a 3-4 year duration. S-PLUS will cover an 
additional $\sim1200\deg^2$ of the Galactic plane and the Bulge. S-PLUS is a Brazilian-led project 
funded by Brazil with important financial support from the Universidad de La Serena in Chile (1 FTE) 
and practical support from CEFCA (Spain).}, the southern counterpart of J-PLUS, while 8 are accessible 
from the northern facility.

Finally, it is interesting to remark that the J-PLUS (and S-PLUS) T80Cam FoV ($1.4\degr\times1.4\degr$) 
is wide enough to collect multi-filter photometry reaching beyond the tidal radius of the vast majority 
of the Galactic GCs, the exceptions being 47\,Tuc, NGC\,6752 and Omega Cen, for which the tidal radius 
is slightly larger than the FoV (\citealt{Harris2010}). Thus, it's clear that these narrow-band surveys 
can provide photometrically uniform and relatively deep observations encompassing most of the member 
stars in a single pointing for all Galactic GCs. In addition, it prepares the community to exploit the 
potential of narrow-band filter diagnostics to identify and characterize MPs. This is timely in light 
of the upcoming J-PAS, when we might be able to analyse the chemistry of the SEDs through its extensive
set of 54 narrow-band filters. This  opens a window through which to study properties - including multiple 
stellar populations - in different spatial regions of a GC body, for clusters characterized by different 
metallicity values and located in different environments. 

\begin{acknowledgements}
The authors thank an anonymous referee for interesting comments and suggestions.
We acknowledge financial support from the Spanish Ministry of Economy and Competitiveness through 
grants AYA2012-30789, AYA2015-66211-C2-1-P and AYA2015-66211-C2-2. CJB, ACS and PC thank CNPq for 
partially financing this work. LSJ acknowledges support from FAPESP (0800) and CNPq. We acknowledge 
the OAJ Data Processing and Archiving Unit (UPAD) for reducing and calibrating the OAJ data used in 
this work. Based on observations made with the JAST/T80 telescope at the Observatorio Astrof\'\i sico 
de Javalambre, in Teruel, owned, managed and operated by the Centro de Estudios de F\'\i sica del Cosmos 
de Arag\'on. We acknowledge the OAJ Data Processing and Archiving Unit (UPAD) for reducing and calibrating 
the OAJ data used in this work. Funding for the J-PLUS Project has been provided by the Governments of 
Spain and Arag\'on through the Fondo de Inversiones de Teruel, the Arag\'on Government through the Reseach 
Groups E96 and E103, the Spanish Ministry of Economy and Competitiveness (MINECO; under grants 
AYA2015-66211-C2-1-P, AYA2015-66211-C2-2, AYA2012-30789 and ICTS-2009-14), and European FEDER funding 
(FCDD10-4E-867, FCDD13-4E-2685). RAD acknowledges support from CNPq through BP grant 312307/2015-2, CSIC through grant COOPB20263, FINEP grants REF. 1217/13 - 01.13.0279.00 and REF 0859/10 - 01.10.0663.00 for partial hardware support for the J-PLUS project through the National Observatory of Brazil.
\end{acknowledgements}

\bibliographystyle{aa} 
\bibliography{M15_ref}

\begin{thebibliography}{47}
\expandafter\ifx\csname natexlab\endcsname\relax\def\natexlab#1{#1}\fi

\bibitem[{{Alessandrini} {et~al.}(2016){Alessandrini}, {Lanzoni}, {Ferraro},
  {Miocchi}, \& {Vesperini}}]{Alessandrini16}
{Alessandrini}, E., {Lanzoni}, B., {Ferraro}, F.~R., {Miocchi}, P., \&
  {Vesperini}, E. 2016, \apj, 833, 252

\bibitem[{{Anderson} {et~al.}(2008){Anderson}, {Sarajedini}, {Bedin}, {King},
  {Piotto}, {Reid}, {Siegel}, {Majewski}, {Paust}, {Aparicio}, {Milone},
  {Chaboyer}, \& {Rosenberg}}]{Anderson2008}
{Anderson}, J., {Sarajedini}, A., {Bedin}, L.~R., {et~al.} 2008, \aj, 135, 2055

\bibitem[{{Bastian} \& {Lardo}(2017)}]{Bastian2017}
{Bastian}, N. \& {Lardo}, C. 2017, ArXiv e-prints [\eprint[arXiv]{1712.01286}]

\bibitem[{{Baumgardt}(2017)}]{Baumgardt2017}
{Baumgardt}, H. 2017, \mnras, 464, 2174

\bibitem[{{Bertin} \& {Arnouts}(1996)}]{Bertin96}
{Bertin}, E. \& {Arnouts}, S. 1996, \aaps, 117, 393

\bibitem[{{Bonatto} \& {Bica}(2007)}]{Bonatto07}
{Bonatto}, C. \& {Bica}, E. 2007, \mnras, 377, 1301

\bibitem[{{Carretta}(2015)}]{Carretta2015}
{Carretta}, E. 2015, \apj, 810, 148

\bibitem[{{Carretta} {et~al.}(2009){Carretta}, {Bragaglia}, {Gratton},
  {D'Orazi}, \& {Lucatello}}]{Carretta2009}
{Carretta}, E., {Bragaglia}, A., {Gratton}, R., {D'Orazi}, V., \& {Lucatello},
  S. 2009, \aap, 508, 695

\bibitem[{{Cassisi} {et~al.}(2017){Cassisi}, {Salaris}, {Pietrinferni}, \&
  {Hyder}}]{Cassisi2017}
{Cassisi}, S., {Salaris}, M., {Pietrinferni}, A., \& {Hyder}, D. 2017, \mnras,
  464, 2341

\bibitem[{{Cenarro} {et~al.}(2018){Cenarro}, {Crist{\'o}bal-Hornillos},
  {Mar{\'{\i}}n-Franch}, \& {J-PLUS collaboration}}]{cenarro18}
{Cenarro}, A.~J., {Crist{\'o}bal-Hornillos}, D., {Mar{\'{\i}}n-Franch}, A., \&
  {J-PLUS collaboration}. 2018, \aap, submitted

\bibitem[{{Coelho} {et~al.}(2012){Coelho}, {Percival}, \&
  {Salaris}}]{coelho+12}
{Coelho}, P., {Percival}, S., \& {Salaris}, M. 2012, in Astronomical Society of
  India Conference Series, Vol.~6, Astronomical Society of India Conference
  Series, ed. P.~{Prugniel} \& H.~P. {Singh}, 107

\bibitem[{{Coelho} {et~al.}(2011){Coelho}, {Percival}, \&
  {Salaris}}]{coelho+11}
{Coelho}, P., {Percival}, S.~M., \& {Salaris}, M. 2011, \apj, 734, 72

\bibitem[{{Coelho}(2014)}]{coelho+14}
{Coelho}, P.~R.~T. 2014, \mnras, 440, 1027

\bibitem[{{Dabringhausen} {et~al.}(2008){Dabringhausen}, {Hilker}, \&
  {Kroupa}}]{Dabringhausen2008}
{Dabringhausen}, J., {Hilker}, M., \& {Kroupa}, P. 2008, \mnras, 386, 864

\bibitem[{{Dalessandro} {et~al.}(2014){Dalessandro}, {Massari}, {Bellazzini},
  {Miocchi}, {Mucciarelli}, {Salaris}, {Cassisi}, {Ferraro}, \&
  {Lanzoni}}]{Dalessandro2014}
{Dalessandro}, E., {Massari}, D., {Bellazzini}, M., {et~al.} 2014, \apjl, 791,
  L4

\bibitem[{{de Marchi} \& {Paresce}(1996)}]{MarchiParesce96}
{de Marchi}, G. \& {Paresce}, F. 1996, \apj, 467, 658

\bibitem[{{Djorgovski} \& {King}(1986)}]{DjorKing1986}
{Djorgovski}, S. \& {King}, I.~R. 1986, \apjl, 305, L61

\bibitem[{{Gratton} {et~al.}(2010){Gratton}, {Carretta}, {Bragaglia},
  {Lucatello}, \& {D'Orazi}}]{Gratton2010}
{Gratton}, R., {Carretta}, E., {Bragaglia}, A., {Lucatello}, S., \& {D'Orazi},
  V. 2010, The Messenger, 142, 28

\bibitem[{{Gratton} {et~al.}(2012){Gratton}, {Carretta}, \&
  {Bragaglia}}]{Gratton2012}
{Gratton}, R.~G., {Carretta}, E., \& {Bragaglia}, A. 2012, \aapr, 20, 50

\bibitem[{{Gruyters} {et~al.}(2017){Gruyters}, {Casagrande}, {Milone},
  {Hodgkin}, {Serenelli}, \& {Feltzing}}]{Gruyters2017}
{Gruyters}, P., {Casagrande}, L., {Milone}, A.~P., {et~al.} 2017, \aap, 603,
  A37

\bibitem[{{Harris}(2010)}]{Harris2010}
{Harris}, W.~E. 2010, ArXiv e-prints [\eprint[arXiv]{1012.3224}]

\bibitem[{{Hess}(1924)}]{Hess24}
{Hess}, R. 1924, Probleme der Astronomie, 265

\bibitem[{{Khalisi} {et~al.}(2007){Khalisi}, {Amaro-Seoane}, \&
  {Spurzem}}]{Khalisi2007}
{Khalisi}, E., {Amaro-Seoane}, P., \& {Spurzem}, R. 2007, \mnras, 374, 703

\bibitem[{{King}(1962)}]{King1962}
{King}, I. 1962, \aj, 67, 471

\bibitem[{{Kirsten} {et~al.}(2014){Kirsten}, {Vlemmings}, {Freire}, {Kramer},
  {Rottmann}, \& {Campbell}}]{Kirsten2014}
{Kirsten}, F., {Vlemmings}, W., {Freire}, P., {et~al.} 2014, \aap, 565, A43

\bibitem[{{Kustner}(1921)}]{Kustner21}
{Kustner}, F. 1921, Veroeffentlichungen des Astronomisches Institute der
  Universitaet Bonn, 15

\bibitem[{{Lardo} {et~al.}(2011){Lardo}, {Bellazzini}, {Pancino}, {Carretta},
  {Bragaglia}, \& {Dalessandro}}]{Lardo2011}
{Lardo}, C., {Bellazzini}, M., {Pancino}, E., {et~al.} 2011, \aap, 525, A114

\bibitem[{{Larsen} {et~al.}(2015){Larsen}, {Baumgardt}, {Bastian}, {Brodie},
  {Grundahl}, \& {Strader}}]{Larsen2015}
{Larsen}, S.~S., {Baumgardt}, H., {Bastian}, N., {et~al.} 2015, \apj, 804, 71

\bibitem[{{Logro{\~n}o-Garc{\'{\i}}a}
  {et~al.}(2018){Logro{\~n}o-Garc{\'{\i}}a}, {Vilella-Rojo},
  {L{\'o}pez-Sanjuan}, \& {J-PLUS collaboration}}]{logronho18}
{Logro{\~n}o-Garc{\'{\i}}a}, R., {Vilella-Rojo}, G., {L{\'o}pez-Sanjuan}, C.,
  \& {J-PLUS collaboration}. 2018, \aap, submitted

\bibitem[{{L{\'o}pez-Sanjuan} {et~al.}(2018){L{\'o}pez-Sanjuan},
  {V{\'a}zquez-Rami{\'o}}, {Varela}, \& {J-PLUS collaboration}}]{clsj18}
{L{\'o}pez-Sanjuan}, C., {V{\'a}zquez-Rami{\'o}}, H., {Varela}, J., \& {J-PLUS
  collaboration}. 2018, \aap, submitted

\bibitem[{{McLaughlin} \& {Fall}(2008)}]{McLFall08}
{McLaughlin}, D.~E. \& {Fall}, S.~M. 2008, \apj, 679, 1272

\bibitem[{{Milone} {et~al.}(2014){Milone}, {Marino}, {Bedin}, {Piotto},
  {Cassisi}, {Dieball}, {Anderson}, {Jerjen}, {Asplund}, {Bellini}, {Brogaard},
  {Dotter}, {Giersz}, {Heggie}, {Knigge}, {Rich}, {van den Berg}, \&
  {Buonanno}}]{Milone2014}
{Milone}, A.~P., {Marino}, A.~F., {Bedin}, L.~R., {et~al.} 2014, \mnras, 439,
  1588

\bibitem[{{Milone} {et~al.}(2017){Milone}, {Marino}, {D'Antona}, {Bedin},
  {Piotto}, {Jerjen}, {Anderson}, {Dotter}, {Criscienzo}, \&
  {Lagioia}}]{Milone2017}
{Milone}, A.~P., {Marino}, A.~F., {D'Antona}, F., {et~al.} 2017, \mnras, 465,
  4363

\bibitem[{{Milone} {et~al.}(2013){Milone}, {Marino}, {Piotto}, {Bedin},
  {Anderson}, {Aparicio}, {Bellini}, {Cassisi}, {D'Antona}, {Grundahl},
  {Monelli}, \& {Yong}}]{Milone2013}
{Milone}, A.~P., {Marino}, A.~F., {Piotto}, G., {et~al.} 2013, \apj, 767, 120

\bibitem[{{Milone} {et~al.}(2009){Milone}, {Stetson}, {Piotto}, {Bedin},
  {Anderson}, {Cassisi}, \& {Salaris}}]{Milone2009}
{Milone}, A.~P., {Stetson}, P.~B., {Piotto}, G., {et~al.} 2009, \aap, 503, 755

\bibitem[{{Miocchi} {et~al.}(2013){Miocchi}, {Lanzoni}, {Ferraro},
  {Dalessandro}, {Vesperini}, {Pasquato}, {Beccari}, {Pallanca}, \&
  {Sanna}}]{Miocchi2013}
{Miocchi}, P., {Lanzoni}, B., {Ferraro}, F.~R., {et~al.} 2013, \apj, 774, 151

\bibitem[{{Molino} {et~al.}(2018){Molino}, {Costa-Duarte}, {Mendes de
  Oliveira}, \& {J-PLUS collaboration}}]{molino18}
{Molino}, A., {Costa-Duarte}, M.~V., {Mendes de Oliveira}, C., \& {J-PLUS
  collaboration}. 2018, \aap, submitted

\bibitem[{{Nardiello} {et~al.}(2015){Nardiello}, {Milone}, {Piotto}, {Marino},
  {Bellini}, \& {Cassisi}}]{Nardiello2015}
{Nardiello}, D., {Milone}, A.~P., {Piotto}, G., {et~al.} 2015, \aap, 573, A70

\bibitem[{{Noyola} \& {Gebhardt}(2007)}]{Noyola2007}
{Noyola}, E. \& {Gebhardt}, K. 2007, \aj, 134, 912

\bibitem[{{Pancino} {et~al.}(2010){Pancino}, {Rejkuba}, {Zoccali}, \&
  {Carrera}}]{Pancino2010}
{Pancino}, E., {Rejkuba}, M., {Zoccali}, M., \& {Carrera}, R. 2010, \aap, 524,
  A44

\bibitem[{{Pease}(1928)}]{Pease28}
{Pease}, F.~G. 1928, \pasp, 40, 342

\bibitem[{{Piotto} {et~al.}(2007){Piotto}, {Bedin}, {Anderson}, {King},
  {Cassisi}, {Milone}, {Villanova}, {Pietrinferni}, \& {Renzini}}]{Piotto2007}
{Piotto}, G., {Bedin}, L.~R., {Anderson}, J., {et~al.} 2007, \apjl, 661, L53

\bibitem[{{Piotto} {et~al.}(2015){Piotto}, {Milone}, {Bedin}, {Anderson},
  {King}, {Marino}, {Nardiello}, {Aparicio}, {Barbuy}, {Bellini}, {Brown},
  {Cassisi}, {Cool}, {Cunial}, {Dalessandro}, {D'Antona}, {Ferraro}, {Hidalgo},
  {Lanzoni}, {Monelli}, {Ortolani}, {Renzini}, {Salaris}, {Sarajedini}, {van
  der Marel}, {Vesperini}, \& {Zoccali}}]{Piotto2015}
{Piotto}, G., {Milone}, A.~P., {Bedin}, L.~R., {et~al.} 2015, \aj, 149, 91

\bibitem[{{San Roman} {et~al.}(2018){San Roman}, {S{\'a}nchez-Bl{\'a}zquez},
  {Cenarro}, \& {J-PLUS collaboration}}]{sanroman18}
{San Roman}, I., {S{\'a}nchez-Bl{\'a}zquez}, P., {Cenarro}, A.~J., \& {J-PLUS
  collaboration}. 2018, \aap, in prep.

\bibitem[{{Sarajedini} {et~al.}(2007){Sarajedini}, {Bedin}, {Chaboyer},
  {Dotter}, {Siegel}, {Anderson}, {Aparicio}, {King}, {Majewski},
  {Mar{\'{\i}}n-Franch}, {Piotto}, {Reid}, \& {Rosenberg}}]{Sarajedini2007}
{Sarajedini}, A., {Bedin}, L.~R., {Chaboyer}, B., {et~al.} 2007, \aj, 133, 1658

\bibitem[{{Stetson}(1987)}]{Stetson1987}
{Stetson}, P.~B. 1987, \pasp, 99, 191

\bibitem[{{Wilson}(1975)}]{Wilson1975}
{Wilson}, C.~P. 1975, \aj, 80, 175

\end{thebibliography}

\end{document}